%% file: paper.tex
\PassOptionsToPackage{table}{xcolor}
\documentclass[leqno,oneeqnum,onefignum,onetabnum,final]{siamart190516}

\input preamble

\input macros

\title{Fast GPU 3D diffeomorphic image registration}

\author{%
Malte Brunn\footnotemark[1]
\and Naveen Himthani\footnotemark[2]
\and George Biros\footnotemark[2]
\and Miriam Mehl\footnotemark[1]
\and Andreas Mang\footnotemark[3]}

\begin{document}
\maketitle

\renewcommand{\thefootnote}{\fnsymbol{footnote}}

\footnotetext[1]{Institute for Parallel and Distributed Systems, University of Stuttgart, Stuttgart 70569 DE, {\tt malte.brunn@ipvs.uni-stuttgart.de}, {\tt miriam.mehl@ipvs.uni-stuttgart.de}}
\footnotetext[2]{Oden Institute of Computational Engineering and Sciences, The University of Texas at Austin, TX 78712, USA, {\tt naveen@ices.utexas.edu}, {\tt gbiros@acm.org}}
\footnotetext[3]{Department of Mathematics, University of Houston, TX 77204, USA, {\tt andreas@math.uh.edu}}

\begin{abstract}
3D image registration is one of the most fundamental and computationally expensive operations in medical image analysis. Here, we present a mixed-precision, Gauss--Newton--Krylov solver for diffeomorphic registration of two images. Our work extends the publicly available \claire{} library to GPU architectures. Despite the importance of image registration, only a few implementations of large deformation diffeomorphic registration packages support GPUs. Our contributions are new algorithms to significantly reduce the run time of the two main computational kernels in \claire{}: calculation of derivatives and scattered-data interpolation. We deploy (i) highly-optimized, mixed-precision GPU-kernels for the evaluation of scattered-data interpolation, (ii) replace Fast-Fourier-Transform (FFT)-based first-order derivatives with optimized 8th-order finite differences, and (iii) compare with state-of-the-art CPU and GPU implementations. As a highlight, we demonstrate that we can register $256^3$ clinical images in less than 6 seconds on a single NVIDIA Tesla V100. This amounts to over 20$\times$ speed-up over the current version of \claire{} and over 30$\times$ speed-up over existing GPU implementations.
\end{abstract}

\section{INTRODUCTION}\label{s:intro}

Image registration (also known as image alignment, warping, or matching) is an important task in medical image analysis~\cite{Sotiras:2013a}. It is used in computer aided diagnosis and clinical population studies. A comprehensive overview can be found in~\cite{Modersitzki:2004a,Modersitzki:2009a,Fischer:2008a,Sotiras:2013a}. The image registration problem is roughly this: Given two images $\sta_0(\x)$ (the template image) and $\sta_1(\x)$ (the reference image; here, $\x \in \Omega \subset \ns{R}^3$), we seek a spatial transformation $\dmap(\x)$ such that the deformed template image $\sta_0(\dmap(\x))$ is similar to $\sta_1(\x)$~\cite{Modersitzki:2004a}. Registration methods can be classified according to the parameterization for $\dmap$. In this paper, we consider methods that belong or are related to large-deformation diffeomorphic metric mapping (\acr{LDDMM})~\cite{Beg:2005a,Younes:2010a}. Such mappings provide maximal flexibility~\cite{Sotiras:2013a}. \lddmm{} maps are expensive to compute since they are infinite-dimensional. Upon discretization, the number of unknowns for $\dmap$ is still in the millions. For example, registering two $256^3$ images requires calculating a $256^3$ resolution stationary velocity field $\vel(\x)\ns{R}^3$ with $\approx$50\,M unknowns. Furthermore, \lddmm{} registration is a highly non-linear and ill-conditioned inverse problem~\cite{Fischer:2008a}. As a result, image registration can take a few minutes on multi-core high-end CPUs. As large clinical, cross-center, population-study workflows require thousands of registrations, reducing the compute time of a single registration to seconds translates to a reduction of clinical study time from weeks to a few days. GPUs with their inherent parallelism and low energy consumption are an attractive choice to achieve this goal. However, despite the need for high-througput computational performance for registration, and the existence of several software libraries for \lddmm{} registration, there is little work on highly optimized GPU implementations (see \secref{s:relatedwork} below).

\subsection{Contributions}
\label{s:contributions}

Based on the open source diffeomorphic image registration framework \claire~\cite{Mang:2015NK,Mang:2016H1,Mang:2016SC,Mang:2018CLAIRE,claire-web}, we introduce a new, optimized, GPU implementation of \lddmm{} registration. The overall mathematical formulation and solution strategy remains unaltered from~\cite{Mang:2018CLAIRE}. We propose several modifications of the \ipoint{differentiation} and \ipoint{interpolation} kernels, which are the main computational kernels in \claire{}. More specifically, our contributions are:
\begin{itemize}
\item {\bf Interpolation:} The first important computational kernel is scattered-data interpolation used for semi-Lagrangian advection. \claire\ originally employed a Lagrange-basis cubic interpolation. We study several alternative methods on GPUs using a combination of pre-filtering, texture, and polynomial interpolation. We study their accuracy and performance using simple performance models and vendor performance profiling tools in~\secref{s:perf}.
\item {\bf Differentiation:} The second important computational kernel is computing derivatives (gradient and divergence) of 3D images (scalar fields). We introduce a mixed-precision implementation using 8\ts{th} order finite-difference (\acr{FD8}) kernels to replace FFT-based spectral derivatives. In particular, we replace all first order derivatives that appear in the partial differential equations (\acr{PDE}) of our optimality systems. Note that FFTs are still retained for higher-order derivatives and their inverse. We discuss this in detail in~\secref{s:perf}.
\item {\bf Evaluation:}  We evaluate the new algorithm on four Magnetic Resonance Imaging \acr{(MRI)} scans and for three different image resolutions. We compare the proposed method with the original \claire{} in~\secref{s:results} as well as with the GPU packages \sw{PyCA}~\cite{pyca-git} and \sw{deformetrica}~\cite{Bone:2018b,deformetrica-git}. We discuss these experiments in detail in~\secref{s:results}. Overall, the method is over $20\times$ faster than the original CPU-based \claire{} and produces registration maps of similar quality. This speedup does not only reflect hardware differences but mostly algorithmic changes, some of which could also be implemented in a CPU version. Furthermore, reducing the accuracy of certain calculations to exploit hardware acceleration has no negative effects on the quality of the registration.
\end{itemize}

\subsection{Limitations}
\label{s:limitations}

The original implementation of \claire{} was built to support the Message Passing Interface (\acr{MPI}) for parallelism~\cite{Mang:2016SC,Mang:2018CLAIRE,Gholami:2017SC}. Our proposed adaption for GPUs has not been integrated with MPI yet. This will be subject to future work, in particular the integration of the high-speed GPU interface \textit{NVLink} in a multi-node multi-GPU context. Thus, our solver does not scale to the image sizes that can be handled by \claire{}. However, this is not an issue for clinical images since typical image sizes fit in a single GPU\footnote{The GPU implementation is for a single GPU only and, therefore, limited by the memory available on the considered card (NVIDIA Tesla V100 in our case). The typical size for clinical images (magnetic resonance imaging) is approximately $256^3$ and fits into memory of a single GPU for the current implementation.}.

\subsection{Related Work}
\label{s:relatedwork}

We refer to~\cite{Sotiras:2013a,Modersitzki:2004a,Modersitzki:2009a,Fischer:2008a} for recent developments in image registration. Surveys of GPU accelerated solvers can be found in~\cite{Fluck:2011a,Shams:2010a,Eklund:2013a}. As mentioned above, this work extends \claire~\cite{Mang:2018CLAIRE,claire-web,Mang:2016H1,Gholami:2017SC}. Popular (in clinical studies) software packages for deformable registration are \sw{IRTK} \cite{Rueckert:1999a}, \sw{elastix} \cite{Klein:2010a}, \sw{NiftyReg} \cite{Modat:2010a}, and \sw{FAIR} \cite{Modersitzki:2009a}. GPU implementations of (low-dimensional) parametric approaches are described in~\cite{Modat:2010a,Shackleford:2010a,Shamonin:2014a,Ellingwood:2016a}. Fast GPU implementations of (high-dimensional) nonparametric formulations available in~\sw{FAIR} are presented in~\cite{Koenig:2018a,Budelmann:2019a}. Unlike \claire{}, these methods do not guarantee that the computed map $\dmap$ is a diffeomorphism. One possibility to safeguard against non-diffeomorphic maps $\dmap$ is by augmenting the formulation by hard and/or soft constraints on $\dmap$~\cite{Burger:2013a}, which introduces significant algorithmic complications. Another approach to enable diffeomorphic registration is to parametrize $\dmap$ via a smooth velocity field $\vel$~\cite{Trouve:1998a,Dupuis:1998a}. This approach has been termed \lddmm. The formulation in \claire{} is closely related to \lddmm. A key difference is that \lddmm\ is based on non-stationary (time-dependent) $\vel$ but \claire\ uses stationary $\vel$. Other approaches that use stationary $\vel$ are described in~\cite{Arsigny:2006a,Ashburner:2007a,Hernandez:2009a,Lorenzi:2013a,Lorenzi:2013b,Vercauteren:2009a}. There exists a large body of literature on \lddmm{}-type approaches that, in many cases, mostly focuses on theoretical considerations~\cite{Younes:2010a,Dupuis:1998a,Miller:2001a,Younes:2009a,Younes:2007a}. There is much less work on the design of efficient solvers; examples are~\cite{Beg:2005a,Ashburner:2007a,Vercauteren:2009a,Azencott:2010a,Avants:2011a,Ashburner:2011a,Zhang:2018a,Zhang:2015b,Polzin:2016a}. Popular software packages for \lddmm{} arediffeomorphic \sw{Demons} \cite{Vercauteren:2009a}, \sw{ANTs} \cite{Avants:2011a,Avants:2008a}, \sw{DARTEL} \cite{Ashburner:2007a}, \sw{deformetrica}~\cite{Bone:2018b,deformetrica-git,Bone:2018a,Fishbaugh:2017a}, and \sw{PyCA}~\cite{pyca-git}. A GPU implementation of the diffeomorphic \sw{Demons} algorithm is described in~\cite{Gu:2009a,Courty:2008a}. The runtime reported in~\cite{Courty:2008a} is in the order of \SI{60}{\second} on a Quadro FX 1400 for a dataset of size $128^3$~\cite{Courty:2008a} (\SI{2}{\second} per iteration)\footnote{All timings here are for single-precision calculations, which is typically used in practice. Our results for the proposed method are for single-precision as well.}. A multi-GPU implementation of \sw{DARTEL} is described in~\cite{ValeroLara:2013a,ValeroLara:2014a}.
The work in~\cite{Zhang:2018a} introduces \sw{FLASH}, a fast CPU implementation for \lddmm. It is based on a band-limited spectral discretization targeting low resolution images to speed up the computations. By truncating the problem to 16 frequencies along each spatial dimension, the runtime is reduced from \SI{45}{\second} to under \SI{2}{\second} per iteration, resulting in an overall execution time of $\approx$\SI{200}{\second} for 100 gradient descent steps. In~\cite{Ha:2009a,Ha:2011a}, a (multi-)GPU implementation of the \lddmm\ approach described in~\cite{Joshi:2005a} is presented; the runtime of this solver is in the order of \SI{12}{\second} on a single NVIDIA Quadro FX5600 for a dataset of size $256^3$~\cite{Ha:2009a}. In~\cite{Grzech:2019a}, a GPU accelerated \lddmm{} implementation called \sw{FastReg} is introduced. The authors report results for neuroimaging data with an average DICE of $\approx$0.67 (much smaller than our results) and a runtime of $\approx$\SI{35}{\second} on a GeForce RTX 2080Ti. A GPU implementation of an \lddmm{} formulation for point cloud matching (not images) is described in~\cite{Sommer:2011a}. The software package \sw{deformetrica}~\cite{Bone:2018b} parametrizes $\dmap$ by a finite set of control points~\cite{Durrleman:2014a}. The gradient is computed via automatic differentiation~\cite{pytorch-git}. The timings reported in~\cite{Bone:2018b} for the registration of an image of size $181 \times 217 \times 181$, executing 50 iterations, are \SI{102}{\second} and \SI{202}{\second} (Nvidia Quadro M4000) for two variants of the GPU implementation, respectively. The execution time for the CPU version of \sw{deformetrica} is $\approx$\SI{10}{\hour} (Intel Xeon E5-1630). The runtime for the GPU variant of \sw{PyCA}~\cite{pyca-git} reported in~\cite{Yang:2017a} for a $229\times193\times193$ neuroimaging dataset is \SI{648}{\second} (Nvidia TitanX (Pascal)). Many of these methods reduce the unknowns by using coarser resolutions, and use algorithms that produce a registration quality that is not as good as \claire{} in terms of Jacobians.

Another approach that can speed up image registration is deep learning~\cite{Yang:2017a,Yang:2016a,Krebs:2019a,Balakrishnan:2019a}. As an example, the training in~\cite{Yang:2016a} is performed with \sw{PyCA}; it takes $\approx$\SI{72}{\hour}.  After training, the reported runtime for the registration of $229 \times 193 \times 193$ images is \SI{18.43}{\second} on a single Nvidia TitanX (Pascal)~\cite{Yang:2017a}, which is significantly slower than our method. Most importantly, it is unclear how deep learning performs on unseen clinical datasets.

\subsection{Outline}

We summarize the overall formulation~\secref{s:formulation} and algorithms~\secref{s:numerics} in \claire{}. All material in \secref{s:formulation} and \secref{s:numerics} is discussed in detail in the works~\cite{Mang:2016H1,Mang:2016SC,Mang:2018CLAIRE,Gholami:2017SC,Mang:2017SL}. In~\secref{s:kernels}, we present the two main computational kernels, the scattered-data interpolation and the approximation of first-order spatial derivatives.

\section{METHODS}\label{s:methods}

\subsection{Formulation}
\label{s:formulation}

\claire{} uses an optimal control formulation. Instead of solving for the \lddmm{} $\dmap(x)$, it reformulates the problem for a \ipoint{velocity} $\vel(\x)$ that generates $\dmap(x)$. Specifically, given two images $\sta_0(\x)$ (template image; image to be registered to reference image) and $\sta_1(\x)$ (reference image), we seek a \emph{stationary} velocity field $\vel(\x)$ by solving
\begin{subequations}
\label{e:problem}
\begin{align}
\label{e:varopt:objective}
\minopt_{\vel} & \; \half{1} \int_{\Omega} (\sta(\x,1) - \sta_1(\x))^2 \dx
 + \frac{\beta}{2} \int_{\Omega} \langle\dop{A}\vel(\x),\vel(\x)\rangle\dx \\
\begin{aligned}
\label{e:varopt:constraint}
\text{subject to}
\\\\
\end{aligned}
& \;\;
\begin{aligned}
\p_t \sta(\x,t) + \vel(\x) \cdot \igrad \sta(\x,t) &= 0 && \text{in}\;\Omega \times(0,1], \\
\sta(\x,t) &= m_0(\x)                                   && \text{in}\;\Omega \times\{0\}
\end{aligned}
\end{align}
\end{subequations}

\noindent with periodic boundary conditions on $\p\Omega$. The PDE constraint in~\eqref{e:varopt:constraint} is the forward problem of our formulation describing the deformation of the state variable $\sta(\x,t)$. Given a candidate $\vel(\x)$, we model the geometric transformation of the template image $\sta_0(\x)$ by transporting its intensities forward in time. The first term in~\eqref{e:varopt:objective} is an image similarity term (without loss of generality, we use the squared $L^2$-distance). The second term in~\eqref{e:varopt:objective} is a Tikhonov regularization functional with regularization parameter $\beta > 0$. It is introduced to ensure smoothness of $\vel(\x)$ so that the geometric transformation of $\sta_0(\x)$ exists and is a diffeomorphism. We refer to~\cite{ Beg:2005a, Younes:2010a, Barbu:2016a, Borzi:2002a, Vialard:2012a, Chen:2011a} for a theoretical discussion about uniqueness and well-posedness of the forward and inverse problem. We follow the default configuration of \claire{} and select $\dop{A}$ to be a vector Laplacian combined with an additional penalty on the divergence of $\vel$. We refer to~\cite{Mang:2018CLAIRE,Mang:2016H1} for details.

\subsection{Discretization and Numerical Algorithms}
\label{s:numerics}

We use a second-order gradient based method to solve the PDE-constrained optimization problem~\eqref{e:problem}.
The gradient is given by the first-order optimality conditions. We use the method of Lagrange multipliers, and take variations with respect to $\sta$, $\adj$ (adjoint variable introduced below), and $\vel$. The first-order optimality conditions amount to a set of coupled, nonlinear, hyperbolic-elliptic PDEs in 4D (space-time). The Lagrangian is given by
\begin{equation*}
\begin{aligned}
&\fun{L}[\sta,\adj,\vel] =
\half{1}\int_{\Omega}(\sta(\x,1) - \sta_1(\x))^2\dx
+ \frac{\beta}{2}\int_{\Omega}\langle\dop{A}\vel(\x),\vel(\x)\rangle\dx\\
&+ \int_0^1 \int_\Omega \adj(\x,t) (\p_t \sta + \vel \cdot \igrad \sta )\dx\dt 
\end{aligned}
\end{equation*}

\subsubsection{Optimality Conditions \& Reduced Space Approach}
\label{s:optcond}

The first order optimality conditions of \eqref{e:problem} consist of three equations. First, the forward problem \eqref{e:varopt:constraint} (variation of $\fun{L}$ with respect to $\adj$). Second,  the backward in time adjoint problem (variation of $\fun{L}$ with respect to $\sta$):
\begin{equation}\label{e:adj-transport}
\begin{aligned}
-\p_t\adj(\x,t) - \idiv \adj(\x,t)\vel(\x) & = 0 && \text{in } \Omega\times [0,1), \\
\adj(\x,t) &=  m_1(\x) - m(\x,t) &&\text{in } \Omega \times\{1\}
\end{aligned}
\end{equation}

\noindent with periodic boundary conditions on $\p\Omega$. Third, the so-called reduced gradient system (variation of $\fun{L}$ with respect to $\vel$) $\vect{g}(\vel) = \vect{0}$, where
\begin{equation}
\label{e:reducedgrad}
\vect{g}(\vel) \defeq \beta \dop{A}\vel(\x) + \int_0^1 \adj(\x,t)\igrad\sta(\x,t)\dt \quad\text{in } \Omega.
\end{equation}

\noindent \claire{} uses a reduced-space approach, i.e., it iterates on the reduced-space  of $\vel$: given the current iterate $\vel(\x)$, it solves for $\sta(\x,t)$ and $\adj(\x,t)$ using~\eqref{e:varopt:constraint} and~\eqref{e:adj-transport}, and substitutes $\sta$ and $\adj$ to evaluate the gradient $\vect{g}(\vel)$. \claire{} uses a Newton--Krylov method to solve the reduced gradient system $\vect{g}(\vel) = \vect{0}$ for $\vel$. We provide more details in \secref{s:newton}.

\subsubsection{Discretization}
\label{s:discretization}

\begin{figure}
 \begin{minipage}[c]{0.45\textwidth}
\centering
\includegraphics[height=3cm]{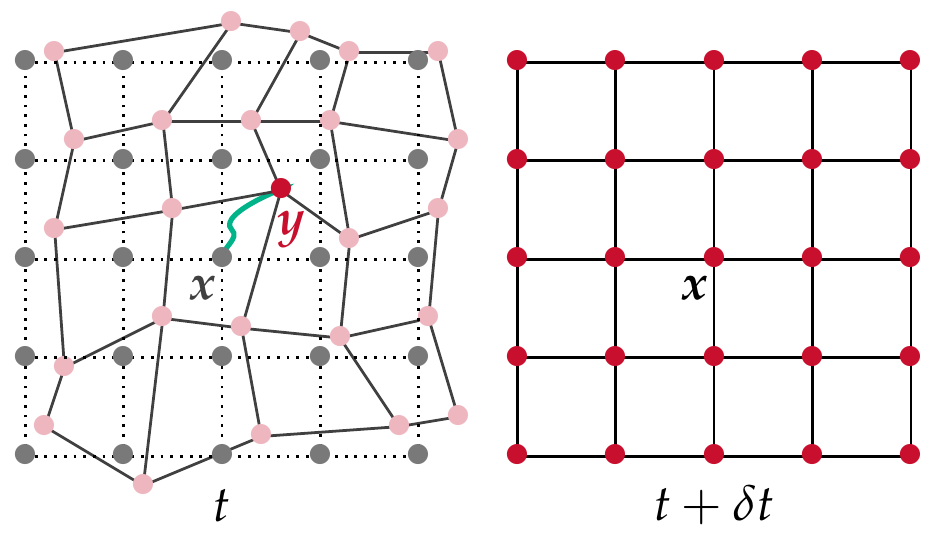}
\end{minipage}
 \begin{minipage}[c]{0.50\textwidth}
\caption{Illustration of the computation of the characteristic in the semi-Lagrangian scheme. We start with a regular grid at time $t+\delta t$ and solve for the characteristic $\vect{y}$ at a given point $\vect{x}$ backward in time (green line in the graphic on the left). The deformed grid configuration is overlaid onto the initial regular grid at time $t$. (Figure modified from \cite{Mang:2017SL}.)}
\label{fig:illustration-sml}
\end{minipage}
\end{figure}

In \claire{}, the forward and the adjoint systems of PDEs \eqref{e:varopt:constraint} and \eqref{e:adj-transport} are discretized in the space-time interval $\Omega \times [0, 1]$, $\Omega \defeq (0, 2 \pi)^3\subset\ns{R}^3$. All spatial fields are periodic in space and discretized using $N = N_1 N_2 N_3$ equispaced grid points $x_{ijk}$. \claire{} uses $N_t$ time steps for the forward and adjoint problems and a semi-Lagrangian scheme (see \figref{fig:illustration-sml}) for the transport equations (see \cite{Mang:2017SL,Mang:2016SC}). It is implemented in two steps: \bipa\item the solution of an ODE $\p_t \traj(t) = \vel(\traj(t))$ in $[t,t+\delta t)$ with final condition $\traj(t+\delta t)=\x$ backward in time to compute the characteristic $\traj$ along which points move; \item the solution of an ODE along this characteristic $\traj$ is used to compute the change of a transported quantity of interest\eipa.

Furthermore, \claire{} uses FFT-based spectral differentiation in several places. The linearized forward problem requires the gradient operator. The adjoint problem requires the computation of the divergence operator. The reduced gradient~\eqref{e:reducedgrad} involves $\dop{A}$ (which is a vector Laplacian), a Leray projection, and a gradient operator (see \cite{Mang:2017SL} for details on the formulation). Spectral differentiation was chosen because it diagonalizes $\dop{A}$. Using a different scheme would introduce significant complications. But the divergence and gradient operators, which are \emph{applied for each time point}, do not need to be done with FFTs, and this is what we exploit in~\secref{s:kernels} to accelerate \claire{}.

\subsubsection{Newton--Krylov Solver}
\label{s:newton}
\claire{} uses a Gauss--Newton--Krylov method globalized with an Armijo line search to find the root of \eqref{e:reducedgrad} for $\vel$. This separates \claire{} from many of the existing registration packages for velocity-based diffeomorphic image registration (see \secref{s:relatedwork} for a discussion). Developing second-order methods for large-scale, nonlinear control problems presents us with numerous challenges \cite{Biros:2005parallel1, Biros:2005parallel2, Boggs:1995a, Hinze:2009a}. If implemented naively, these methods can become computationally prohibitive, despite their improved rate of convergence.

We iterate on the discretized velocity $\dvel\in\ns{R}^{3N}$ according to
\begin{equation}
\label{e:iter}
\dvel_{k+1} = \dvel_k + \alpha_k \divel_k,
\quad \di{H}\divel_k = -\di{g}_k,
\quad k = 0,1,2,\ldots
\end{equation}

\noindent where $\di{H}\in\ns{R}^{3N,3N}$ is the discretized Gauss-Newton Hessian operator (or simply Hessian for the rest of the paper), $\divel_k\in\ns{R}^{3N}$ is the search direction, $\di{g}_k\in\ns{R}^{3N}$ is the discretized gradient given by~\eqref{e:reducedgrad}, $\alpha_k>0$ is a line search parameter, and $k\in\ns{N}$ is the Gauss--Newton iteration count. To compute $\divel_k$ we have to solve the linear system in~\eqref{e:iter} at each Gauss--Newton step. We cannot form or store $\di{H}$ since it is a $3(N_1N_2N_3)$-by-$3(N_1N_2N_3)$ matrix. We invert $\di{H}$ iteratively using the preconditioned conjugate gradient method (\acr{PCG})~\cite{Hestenes:1952a}. Applying the Hessian to a vector (we refer to this operation as the \emph{Hessian matvec}) is similar to evaluating the gradient in~\eqref{e:reducedgrad}; it requires the solution of two PDEs, one forward in time, and one backward in time. We can split the Hessian operator into two terms, $\di{H} = \di{A} + \di{\tilde{H}}$, where $\di{A} \in\ns{R}^{3N,3N}$ is the discretized regularization operator $\dop{A}$; $\di{\tilde{H}} \in\ns{R}^{3N,3N}$ involves inverses of the state and adjoint operators computed by solving two transport equations. Solving these two PDEs is costly; approximating $\di{H}^{-1}$ using PCG at every Gauss-Newton step takes over 90\% of the runtime of \claire{} for clinical images
~\cite{Mang:2018CLAIRE}.

\subsection{Computational Kernels}
\label{s:kernels}

Let us first summarize the overall algorithm. As we just discussed, we use a Gauss--Newton--Krylov method~\eqref{e:iter} to solve the reduced gradient system $\vect{g}(\vel) = \vect{0}$ for $\vel$. The matrix-free Gauss--Newton Hessian involves solving forward and adjoint hyperbolic PDEs for the linearized \eqref{e:varopt:constraint} and~\eqref{e:adj-transport}. If we use $N_t$ time steps, each Hessian matvec requires $2N_t$ semi-Lagrangian steps, $2 N_t$ gradient operators, and $N_t$ divergence operators. In addition, the Hessian matvec needs $\di{A}$ and its inverse, which are computed as spectral operators using FFTs. All these operators have $\bigO(N)$ complexity per time step, up to a logarithmic prefactor. The total number of Hessian matvecs is the sum of PCG iterations across Newton steps. Table \ref{tab:claire} lists the number of FFTs and interpolations in more detail. The overall method is outlined in Algorithm \ref{alg:claire}. The original \claire{} implementation for CPUs used FFTs for gradients, divergences, $\di{A}$ and $\di{A}^{-1}$, and a highly optimized cubic Lagrange interpolation for the semi-Lagrangian method~\cite{Mang:2018CLAIRE}. We transformed all computational kernels to GPU architectures, and most importantly, we introduced several algorithmic innovations to speed-up both derivatives and interpolations. First, we discuss several options for the interpolation. Second, we replace all gradient and divergence operators with high-order finite-difference (\textbf{FD}) operators.\footnote{We note that low-order (first and second order) FD (and finite volume) operators are a common choice in image registration~\cite{Modersitzki:2004a,Modersitzki:2009a}.} Notice that we keep the spectral differentiation for high-order differential operators, since we need to evaluate their inverses in our solver (spectral preconditioner and Leray projection). Computing their inverses can be done efficiently in the spectral domain; for FD it would require linear solves. We show that, for the given image resolution and floating point accuracy, replacing the spectral methods with high-order FD discretizations allows us to maintain accuracy but significantly increase efficiency on GPUs. To the best of our knowledge, we are the first group to implement this type of mixed-precision code in a hardware and resolution adaptive way. Again, the spectral differentiation is kept for evaluating $\di{A}$ (and its inverse to avoid an additional need for linear solvers); the GPU implementation of the proposed method employs a hybrid differentiation scheme that uses both FFTs and finite differences.

\begin{table}
\caption{We report the complexity of our solver for the compressible case. We report the number of FFT operators (\#FFTs, split into first order derivatives and other, i.e., higher order or inverse operators) and the number of scattered data interpolations (\#IPs) that need to be performed for evaluating the objective functional, the gradient (notice, that the evaluation of the gradient requires forward and adjoint PDE solves), and the Hessian matvec (Gauss--Newton approximation; requires the evaluation of the incremental adjoint and state equations as subfunctions). The first order operators are either implemented as FFT or finite differences (\#FD). We report generic numbers; $d\in\{2,3\}$ denotes the dimension of the ambient space ($d=3$ in our case) and $N_t$ is the number of time steps (we set $N_t=4$). Each Newton iteration requires the evaluation of the objective and the evaluation of the gradient. Each line search step requires the evaluation of the objective function. We demonstrated in~\cite{Mang:2016SC,Mang:2018CLAIRE,Gholami:2017SC} (CPU implementation of CLAIRE) that about 90\% of the runtime is spent on evaluating FFTs and the IP model. To reduce the memory footprint of our solver, we evaluate parts of the gradient and Hessian matvec during the solution of the adjoint operators. The memory pressure is  $\mathcal{O}((N_t + 7)N_1N_2N_3)$ for the gradient and  $\mathcal{O}((N_t + 10)N_1N_2N_3)$ for the Hessian matvec, respectively.}\label{tab:claire}
\centering\scriptsize\renewcommand{\arraystretch}{0.8}
\begin{tabular}{llllll}\toprule
function         & subfunction & symbol      & \#FFTs / \#FD & \#FFTs & \#IPs\\
        &  & symbol      & (1st order) & (other) & \\\midrule
objective functional  & & ---                  & ---            & $d$ & $d + N_t$         \\
& state equation (SE)   & $m$                  & ---            & --- & $d + N_t$         \\ \midrule
gradient              & & $\vect{g}$           & $d(N_t + 2)  $ & $d$ & $d+N_t+1$         \\
& adjoint equation (AE) & $\lambda$            & $d$            & --- & $d + N_t + 1$     \\ \midrule
Hessian matvec        & & $\di{H}\tilde{\vel}$ & $d(2N_t+3)$    & $d$ & $d+(d+2)N_t + 1$  \\
& incremental SE        & $\tilde{m}$          & $d(N_t+1)$     & --- & $d + (d+1)N_t$    \\
& incremental AE        & $\tilde{\lambda}$    & $d$            & --- & $N_t + 1$         \\\bottomrule
\end{tabular}
\end{table}

\begin{algorithm}
\caption{Basic algorithm for a Gauss--Newton-Krylov step~\eqref{e:iter} in \claire{} to solve the reduced gradient system $\vect{g}(\vel) = \vect{0}$ for $\vel$.}\label{alg:claire}
\SetKwFor{Run}{loop}{}{}
\SetKwFor{Block}{}{}{}
\SetKw{Line}{\ }
\Run(\algcomment{Newton method ($\mathbf{g}(\vel) = 0$)}){root of \eqref{e:reducedgrad}}{
\Block{$\textsc{ObjectiveFunctional}(\vel)$\algcomment{as defined in \eqref{e:varopt:objective}}}{
\Line{$m \gets \textsc{StateEquation}(\vel, m_0)$}\algcomment{\eqref{e:varopt:constraint}}\\
}
\Block{$\textsc{Gradient}(\vel)$\algcomment{\eqref{e:reducedgrad}}}{
\Line{$\lambda \gets \textsc{AdjointEquation}(\vel, m, m_R)$}\algcomment{\eqref{e:adj-transport}}\\
}
\Run(\algcomment{solve \eqref{e:iter}}
){$\textsc{KrylovSolver}(\tilde{\vel}, \epsilon_{K})$}{
\Block{$\textsc{HessianMatVec}(\tilde{\vel})$\algcomment{$\beta\mathcal{A}\tilde{\vel} + \int\tilde\lambda\nabla m~\mathrm{d}t$}}{
\Line{$\tilde m \gets \textsc{IncStateEquation}(\vel, \tilde{\vel})$}\algcomment{$\partial_t \tilde m + \vel\cdot\nabla\tilde m + \tilde{\vel}\cdot\nabla m=0$}\\
\Line{$\tilde\lambda \gets \textsc{IncAdjointEquation}(\vel, \tilde m)$}\algcomment{$-\partial_t \tilde\lambda - \nabla\cdot\vel\tilde\lambda=0$}\\
}
\Block{$\textsc{Preconditioner}(\mathbf{r})$}{
\Line{$\beta^{-1}\mathcal{A}^{-1}\bvec{r}$}\\
}
}
\Run{$\textsc{LineSearch}(\alpha)$}{
\Block{$\textsc{ObjectiveFunctional}(\vel + \alpha\tilde{\vel})$}{
\Line{$m \gets \textsc{StateEquation}(\vel+\alpha\tilde{\vel}, m_0)$}\\
}
}
\Line{$\vel \gets \vel + \alpha \tilde{\vel}$}\algcomment{Newton step}
}
\end{algorithm}

\subsubsection{GPU Interpolation}

The semi-Lagrangian scheme requires costly interpolation of velocities and scalar image fields along backward characteristics as shown in \figref{fig:illustration-sml}. \claire{} uses Lagrange-based cubic interpolation. GPUs provide two technologies that we exploit in our schemes: texture fetches and hardware support for trilinear interpolation (although not fully single-precision). In addition to these modifications, we also consider another change: switching from Lagrange cubic to B-spline cubic interpolation. The generic formula for interpolating at an off-grid point $\x \defeq (x_1,x_2,x_3)\in\ns{R}^3$ is given by
\begin{align}
f(x_1,x_2,x_3) &= \sum_{i,j,k=0}^{d}c_{ijk}\phi_i(x_1)\phi_j(x_2)\phi_k(x_3),
\label{e:interp}
\end{align}

\noindent where $c_{ijk} \in \ns{R}$ are scalar coefficients associated with each grid point, $d\in\ns{N}$ is the polynomial order, and $\phi_i(x_1)$, $\phi_j(x_2)$, $\phi_k(x_3)$ are the basis functions. For Lagrange interpolation, the coefficients equal the grid values ($c_{ijk}=f_{ijk}$), and the $\phi$'s are the Lagrange polynomials. We use third order cubic ($d=3$) but we also consider first-order trilinear interpolation ($d=1$) since GPUs offer hardware acceleration for it. So, we need to evaluate a set of 64 (cubic) or 8 (linear) grid values $f_{ijk}$. However, there are other options. For example, we can use uniform B-splines for $\phi$. In that case, the coefficients $c_{ijk}$ are non-local---they depend on all grid values $f_{ijk}$ unlike the Lagrange case~\cite{Thevenaz:2012a}.
Below we give the implementation details for the different schemes.
\begin{itemize}
\item {\bf GPU-TXTLIN:} Here we use NVIDIA's libraries for trilinear interpolation\cite{gpugems,cuda-docs}. It is efficiently performed using NVIDIA's hardware-accelerated texture units (using the {\tt tex3D()} function). The texture units store the coefficients of the trilinear interpolation in 9-bit precision and return the result in single precision. We observed some effects in the registration quality in terms of smoothness of the deformation and the overall mismatch---especially in lower-resolutions or when the image has high frequency components.
\item {\bf GPU-LAG:} This is our baseline since it represents a direct translation of the existing algorithm in \claire{} to GPUs. The $c_{ijk}$ values required to evaluate $f$ are ordered lexicographically. This ordering results in non-coalesced memory accesses that reduce performance. To partially improve this, we use the texture function \texttt{tex3D()} as a table lookup to access $c_{ijk}$ and evaluate \eqref{e:interp}. We remark that we use the texture memory only for look ups and not for trilinear interpolation.
\item {\bf GPU-TXTLAG:} This is also a cubic Lagrange interpolation but now we use texture-based interpolation (as opposed to using textures as a table lookup), and thus the accuracy is reduced compared to GPU-LAG. However, in our experiments we don't observe any significant difference in the accuracy. The algorithm is based on the same principle as presented in~\cite{Ruijters:2008a}. Instead of doing \emph{eight} weighted trilinear interpolations, we do \emph{27 weighted trilinear} interpolations at off-grid points. The different number of trilinear interpolations arises due to differences in the Lagrange and B-spline polynomials. Nevertheless, because of hardware acceleration, GPU-TXTLAG significantly outperforms GPU-LAG.
\item {\bf GPU-TXTSPL:} The algorithm we use is exactly the one presented in~\cite{Ruijters:2008a}. The implementation is based on the open source library~\cite{cubictextureinterp-git}, with a major modification related to pre-filtering. We replaced the pre-filter in~\cite{cubictextureinterp-git} with a finite convolution inspired by~\cite{Champagnat2012a}. The pre-filtering to compute the coefficients $c_{ijk}$ then becomes a 15-point axis aligned stencil operation on $f_{ijk}$ and is implemented using the FD scheme used in the CUDA SDK example~\cite{cudafinitedifference-git}. We also modified the code to support periodic boundary conditions. Then, following~\cite{Ruijters:2008a}, we use \emph{eight} weighted trilinear ($8\times 8f_{ijk}$) interpolations to compose the cubic B-spline interpolation. These interpolations require \emph{eight} texture fetches at off-grid points. Overall,  GPU-TXTSPL significantly outperforms GPU-TXTLAG.
\end{itemize}

\subsubsection{GPU Derivatives}
\label{s:derivatives}

The CPU \claire{} uses FFTs to perform spatial differentiation~\cite{Gholami:2017SC}. Since our functions are periodic, all such operators are diagonal in the spectral domain.  But in the proposed GPU implementation, we use an FD scheme that is more accurate (only for the given resolutions---not asymptotically) and faster than FFTs (see~\secref{s:perf}).
\begin{itemize}
\item {\bf Finite Difference Scheme:} In particular, we use an 8\textsuperscript{th} order central difference scheme to evaluate first-order partial derivatives for the gradient and divergence operators. To evaluate the partial derivative at a regular grid point, we require nine axis-aligned function evaluations $f_{ijk}$. We load the grid values $f_{ijk}$ from global memory to a shared memory tile and then evaluate the finite difference  stencil. The derivative evaluations in the $x_1$, $x_2$ and $x_3$ spatial dimensions are independent of each other. Our implementation is the same as the CUDA SDK finite difference code \cite{cudafinitedifference-git} except that our implementation works for general grid sizes and supports periodic boundary conditions.
\item {\bf FFT (Spectral Differentiation):} \claire{} uses \texttt{AccFFT} \cite{accfft_github, accfft-home-page}, which supports MPI for both CPU and GPUs. Here, we just use \texttt{cuFFT} \cite{Nvidia2007b}  as we focus on a single GPU implementation. When we use FFTs for gradient and divergence operations we compute 3D FFTs. This avoids an explicit transpose operation on the data and misaligned memory accesses. Additionally, 3D FFTs reduce the number of memory accesses of the spectral data from global device memory. For the gradient all partial derivatives can be computed with only a single read and three write operation per element (instead of $3+3$ as for one-dimensional FFTs). Similarly, the divergence operator only needs a single store operation after summing all partial derivatives.
\end{itemize}

\section{KERNEL PERFORMANCE ANALYSIS}\label{s:perf}

In this section, we evaluate the performance of interpolation (\acr{IP}) and finite difference (\acr{FD}) kernels. We calculate their arithmetic intensity (or simply \iquote{intensity}) defined as the ratio of FLOPS (total floating point operations) to MOPS (total memory operations). We compare the kernel intensity to the device intensity. If the kernel intensity  is less than the device intensity (peak floating point performance divided by peak device memory bandwidth), then the kernel is memory bound, otherwise it is compute bound. This is a simplification of the roofline model \cite{roofline} since here we do not account for the cache hierarchy and latency effects. We also perform benchmark experiments to identify performance ceilings for our kernels.

As reference system for the CPU code, we used a two-socket Intel Skylake system. It is equipped with two Xeon Gold 5120 with a maximum frequency of $\SI{2.20}{\giga\hertz}$ and a maximum bandwidth of $\SI{107.3}{\giga\byte\per\second}$ with a TDP of $\SI{105}{\watt}$ per socket. We used a 32GB NVidia Tesla V100 with a memory bandwidth $B_{\text{max}}$ of $\SI{900}{\giga\byte\per\second}$ and a TDP of $\SI{300}{\watt}$ for GPU experiments. The V100 is part of a two socket IBM Power9 system featuring NVLink as inter-device bus. Our implementation is in C\texttt{++} and CUDA, and uses the PETSc library~\cite{Balay:2016a} for the Gauss--Newton--Krylov solvers.

\subsection{Cubic Interpolation Kernel}

Both cubic and linear IP are memory bound. The IP kernel has two main inputs: the target point coordinates ($3N$ floats), and the grid point scalar values ($N$ floats). The output is the scalar field at the target points ($N$ floats). Thus, the total MOPS is five floats ($20$\,B) per target point. Formula~\eqref{e:interp} applies to both B-spline and Lagrange interpolation: the value at each target point depends on $64$ regular grid values for cubic and $8$ for trilinear interpolation, and these are not contiguous in memory.

Assuming an infinite amount of fast memory and ignoring latency cost, an analytic calculation of the FLOPS for each kernel gives the arithmetic intensity that shows that the kernels are memory bound. We overestimate the analytic intensity because we assume that all $c_{ijk}$ values in~\eqref{e:interp} are loaded exactly once from device memory, which will typically not be the case, unless the memory accesses are fully coalesced.
We evaluate performance using an effective bandwidth in GB/s defined as $\frac{(b_w + b_r)}{t x 10^9}$, where  $b_r$ and $b_w$ are the kernel loads/stores in bytes and $t$ is the kernel total run time. We tuned the threadblock configuration to obtain optimal performance for the interpolation kernel. We used a one dimensional threadblock configuration with 256 threads for all our experiments. We perform two experiments for a localized and for a scattered target point distribution.

\begin{table*}
\caption{Experiment 2: Comparison of arithmetic \iquote{intensity} for two interpolations with $N=256^3$ on an NVIDIA Tesla V100. For the analytic \iquote{FLOPS} value, we assume that each FPADD (add), FPMUL (multiply), FPSP (other ops like division) is one FLOP, and an FMA (multiply add) is two FLOPS. For GPU-TXTSPL, GPU-TXTLIN and GPU-TXTLAG, the FLOP count includes the operations required to compute the trilinear interpolation done internally by the texture unit. For the analytic \iquote{MOPS},  we assume that each $f_{ijk}$ value is loaded only once from the device memory. (Thus, all kernels have the same MOPS since the fact that we use linear versus cubic doesn't matter for this simple model.)  The intensity value is computed as the ratio of FLOPS/MOPS. For the experimental ``FLOPS'', we make the same assumption as for the analytic ``FLOPS'', but here the FLOP count is obtained from the NVidia Visual Profiler. The experimental ``MOPS'' are also obtained from the visual profiler and are the sum of the total number of bytes read from and written to the GPU device memory by the L2 cache. GPU-TXTSPL$^\star$ corresponds to  GPU-TXTSPL w/o prefilter.}
\label{tab:flopsmops}
\renewcommand{\arraystretch}{0.8}
\centering
\scriptsize
\begin{tabular}{lrrRrrRL}\toprule
\multicolumn{1}{c}{}                         &
\multicolumn{3}{c|}{Analytic} & \multicolumn{3}{c}{Experimental} &
\multicolumn{1}{c}{} \\ \midrule
\textbf{Kernel}    & {\bf FLOPS} & {\bf MOPS} & {\bf intensity} & {\bf GFLOPS} & {\bf GMOPS} & {\bf intensity}& {\bf bound by}\\\midrule
PRE-FILTER         & \num{ 22}   & \num{ 8}   & \num{2.75}      & \num{0.37}   & \num{0.14}          & \num{2.64}    & memory         \\
GPU-TXTLIN         & \num{ 30}   & \num{20}   & \num{1.50}      & \num{0.10}   & \num{0.34}          & \num{0.30}    & memory         \\
GPU-LAG            & \num{221}   & \num{20}   & \num{11.05}     & \num{3.66}   & \num{1.55}          & \num{2.36}    & memory         \\
GPU-TXTLAG         & \num{482}   & \num{20}   & \num{24.1}      & \num{3.00}   & \num{0.34}          & \num{8.94}    & memory         \\
GPU-TXTSPL$^\star$ & \num{294}   & \num{20}   & \num{14.7}      & \num{2.97}   & \num{0.27}          & \num{10.86}   & memory         \\\midrule
\rowcolor{colorA}NVIDIA Tesla V100  &             &            &               & \num{14000}GFLOPS/s & \num{900}GB/s & \num{15.56}    &                \\
\bottomrule
\end{tabular}
\end{table*}

\begin{table*}
\caption{Performance of the overall semi-Lagrangian transport using different interpolation kernels on the V100. We report runtimes (in seconds) for applying an \lddmm{} transformation on a real 3D brain MR image using a semi-Lagrangian scheme. We deform the brain image using a velocity field (generated by registering two images from a clinical dataset) forward in time, followed by deforming the resulting image backward in time. We then compare the original image to the resulting image and compute the relative mismatch between the two. CPU-LAG, GPU-LAG and GPU-TXTLAG have a relative error of \scnum{5.25e-2} and \scnum{2.36e-2} for $N=64^3$ and $256^3$, respectively. GPU-TXTSPL is 2$\times$ more accurate, and has a relative error \scnum{2.50e-2} and \scnum{1.66e-2}, respectively. GPU-TXTLIN has a relative error of \scnum{1.21e-1} and \scnum{5.54e-2}, respectively. We also report wall-clock time for two advection solves, which incurs 14 interpolation kernel calls in total. The corresponding effective global memory bandwidth is also reported. The run time and bandwidth reported for GPU-TXTSPL include the overhead of the pre-filter operation. The CPU Lagrange (CPU-LAG) interpolation kernel is executed on a single intel-skylake node with 24 MPI tasks.}
\label{tab:smlbenchmark}
\centering\scriptsize\renewcommand{\arraystretch}{0.8}
\begin{tabular}{llLlLlLlLlLlL}\toprule
& \multicolumn{1}{c}{CPU-LAG} & \multicolumn{2}{c}{GPU-LAG} &
\multicolumn{2}{c}{GPU-TXTLAG} & \multicolumn{2}{c}{GPU-TXTSPL (w/pre-filter)} & \multicolumn{2}{c}{GPU-TXTLIN} \\ \midrule
$N$     & {\bf time}  & {\bf time}      & {\bf BW} & {\bf time}      & {\bf BW}  & {\bf time}      & {\bf BW}  & {\bf time}      & {\bf BW}  \\ \midrule
$64^3$  &  \num{16  } & \scnum{1.48e0}  & \num{50} & \scnum{6.36e-1} & \num{115} & \scnum{6.72e-1} & \num{240} & \scnum{1.33e-1} & \num{552} \\
$128^3$ &  \num{124 } & \scnum{1.09e1}  & \num{54} & \scnum{4.03e0 } & \num{146} & \scnum{2.92e0 } & \num{442} & \scnum{8.33e-1} & \num{705} \\
$256^3$ &  \num{1000} & \scnum{8.41e1}  & \num{56} & \scnum{3.45e1 } & \num{136} & \scnum{2.24e1 } & \num{461} & \scnum{5.95e0 } & \num{790} \\
\bottomrule
\end{tabular}
\end{table*}

\begin{table}
\caption{Experiment 2: Runtime (in seconds) and error of different interpolation kernels on the NVIDIA Tesla V100. We report the relative interpolation error and the averaged run time for one kernel call in seconds. The relative interpolation error is given in the $\ell^2$-norm with respect to an analytically known function. The evaluation is done on a grid with randomly perturbed grid points. The interpolated function is given by $(sin^2(8 x_1) + sin^2(2 x_2) + sin^2(4 x_3))/3$. For this synthetic setup, the measured runtime $t_\text{syn}$ (syn) is averaged over 100 interpolations. The faster variants GPU-TXTSPL and GPU-TXTLIN were also applied to the real data experiments shown in \secref{s:results}. For those, we also report the per-call duration $t_\text{reg}$ for averaged over all Gauss-Newton iterations. The reported runtimes include all pre- and post-processing needed for the interpolation method.\label{tab:interpolation}}
\centering\scriptsize\renewcommand{\arraystretch}{0.8}
\begin{tabular}{lllll}
\toprule
$N$ & {\bf method} & {\bf error} & {$\mathbf{t_{syn}}$} & {$\mathbf{t_{reg}}$} \\
\midrule
\multirow{4}{*}{$64^3$}
& GPU-LAG    & \scnum{9.851e-3} & \scnum{1.207e-4} & --- \\
& GPU-TXTLAG & \scnum{9.845e-3} & \scnum{7.488e-5} & --- \\
& GPU-TXTSPL & \scnum{2.249e-3} & \scnum{1.140e-4} & \scnum{1.087e-04} \\
& GPU-TXTLIN & \scnum{2.605e-2} & \scnum{3.784e-5} & \scnum{2.710e-05} \\
\midrule
\multirow{4}{*}{$128^3$}
& GPU-LAG    & \scnum{7.193e-4} & \scnum{7.427e-4} & --- \\
& GPU-TXTLAG & \scnum{7.330e-4} & \scnum{4.091e-4} & --- \\
& GPU-TXTSPL & \scnum{1.134e-4} & \scnum{3.602e-4} & \scnum{3.283e-04} \\
& GPU-TXTLIN & \scnum{6.765e-3} & \scnum{1.318e-4} & \scnum{1.376e-04} \\
\midrule
\multirow{4}{*}{$256^3$}
& GPU-LAG    & \scnum{4.670e-5} & \scnum{5.240e-3} & --- \\
& GPU-TXTLAG & \scnum{8.728e-5} & \scnum{3.022e-3} & --- \\
& GPU-TXTSPL & \scnum{4.969e-5} & \scnum{2.324e-3} & \scnum{2.132e-03} \\
& GPU-TXTLIN & \scnum{1.709e-3} & \scnum{8.365e-4} & \scnum{1.024e-03} \\
\bottomrule
\end{tabular}
\end{table}

\subsubsection{Experiment 1---Localized target points}

As we discussed, each target point requires a set of $c_{ijk}$ values. To isolate the memory issues related to streaming the target points, we conducted a run in which \emph{all target points use the same 64 grid values} for interpolation. This ensures full reuse of regular grid values among targets and provides an upper limit for the performance of the kernel. We run this test on the GPU-LAG and GPU-TXTSPL kernels. The performance of GPU-TXTLAG is somewhere in between and we omitted it in these runs. In this model the MOPS change. We only read and write $4N$ floats, and read $64$ grid values for all points. Since all thread blocks need to read these values, the number of total MOPS (in bytes) is equal to $4*(4N + 64*\#\mathrm{threadblocks})$. We use this to estimate an upper performance bound. It is important to note here that the number of threadblocks only matters for a theoretical estimate without accounting for cache effects. In experimental runs, since all threadblocks are accessing the same set of 64 grid values, they will be cached. Hence, different threadblock configurations will not significantly affect the kernel performance, except for extremely small threadblocks where latency effects are dominant.
\begin{itemize}
\item {\bf GPU-LAG Kernel (w/shared memory):} All CUDA thread-blocks load the same set of 64 $c_{ijk}$ values from device memory and store them in the on-chip shared memory for reuse. All threads evaluate the result at their corresponding target points using the data available in shared memory and then apply~\eqref{e:interp}. Using the MOPS estimate from above (and the observed timings), we achieve an effective bandwidth of 570 GB/s (63.3$\%B_{\text{max}}$).
\item {\bf GPU-TXTSPL Kernel:} To calculate the effective bandwidth, we assume that each of the $64$ $c_{ijk}$ are fetched by the texture exactly once from the device memory. Using this assumption (and the observed timings), the effective bandwidth for this method is $350$\,GB/s
(39$\%B_{\mathrm{max}}$). Note that the reported bandwidth here does not account for the prefilter operation. Also note that, in reality, textures cannot take significant advantage of the fact that the target points have exactly the same regular grid dependencies. As a result, there are more memory dependencies (than our MOPS estimate) and, thus, the observed performance drops---compared to the GPU-LAG kernel.
\end{itemize}

\subsubsection{Experiment 2---Scattered target points}

We consider a real distribution (generated via random perturbation of grid points or actual trajectory backward tracking) of target points (and switch to the original 20\,B/point MOPS model). Here, unlike ``Experiment 1'', the implementation of GPU-LAG does not use shared memory to load target point dependencies. The implementation of GPU-LAG which uses shared memory to load target point dependencies for the scattered case is future work. However, the implementation of GPU-TXTSPL remains the same as in ``Experiment 1''.

The analytic observation that the interpolation is memory bound result is confirmed by measurements with the NVIDIA Visual Profiler summarized in  \tabref{tab:flopsmops}.

For a random distribution of target points, GPU-TXTSPL achieves an effective global memory bandwidth of 335 GB/s (37.6$\%B_{\mathrm{max}}$), which is nearly identical to ``Experiment 1''. Hence, GPU-TXTSPL is insensitive to target point dependencies. In contrast, GPU-LAGs performance drops by a factor of 10 to 56 GB/s because we are no longer making explicit use of shared memory to load and reuse the target point dependencies. Also note that, once we coupled the GPU-TXTSPL to the overall semi-Lagrangian scheme in~\tabref{tab:smlbenchmark}, the effective bandwidth increases to 461\,GB/s, which is slightly over 50\% relative to the peak bandwidth. Finally, in~\tabref{tab:interpolation}, we compare the accuracy and time of the four different methods. The differences in accuracy are somewhat significant only in lower resolutions. Note that we get different accuracy results for real brain MR images in \tabref{tab:smlbenchmark}. This is expected since the cubic spline interpolation of GPU-TXTSPL gives better interpolation accuracy than third order Lagrange polynomials used in CPU-LAG or GPU-LAG in cases where the image resolution is not sufficiently high relative to the highest frequency in the image. For the synthetic low frequency image lower used in \tabref{tab:interpolation}, Lagrange polynomials to perform better for higher image resolutions. Here, GPU-LAG gives more accurate results than GPU-TXTSPL for a $256^3$ resolution. We compare our new GPU implementation to the original MPI based CPU version of CLAIRE~\cite{Mang:2018CLAIRE}; the CPU version of CLAIRE does not support OpenMP.

As a byproduct, this analysis also addresses to some extent the following question: would it make sense to reorder (say in Morton order) the target and grid points in order to achieve better locality (but possible sacrifice texture memory)? As we show, an ideal ordering would result in 570\,GB/s; we observe about 460\,GB/s for GPU-TXTSPL and conclude that our implementation is nearly optimal.

\begin{figure}
\centering
\resizebox{0.75\linewidth}{!}{\input{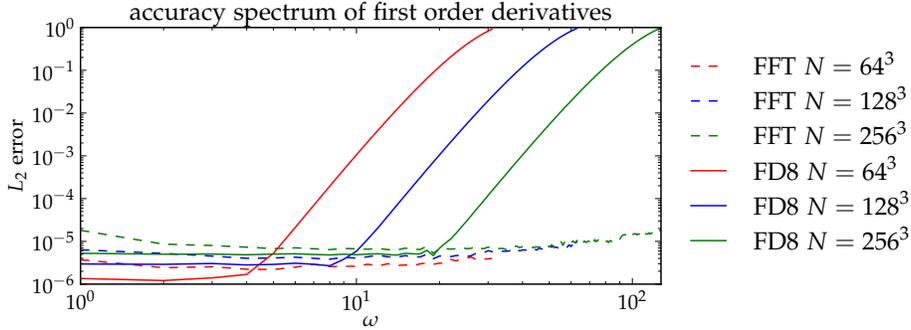}}
\caption{Accuracy of first order differential operators (gradient and divergence) using FFT and 8\textsuperscript{th} order finite differences on a Nvidia Tesla V100 for different problem sizes. We report the $L_2$ error of our operators. The error is measured using the computed partial derivative in $x_3$-direction of the function $\sin(\omega x_3) + \cos(\omega x_3)$ compared to the analytical derivative. The error is plotted over the frequency up to the Nyquist frequency. Finite differences are more accurate for low frequency modes and have an increasing error for higher modes. By replacing FFTs with finite differences, we trade faster computation (due to a higher data locality and a reduced algorithmic complexity) against lower accuracy for high frequency modes.\label{fig:diff-perf}}
\end{figure}

\subsection{Finite Difference Kernel}

In our implementation, each CUDA thread block evaluates the derivatives for a 2D tile of data. We refer to the points contained in this tile as \textit{inner points}. To evaluate the derivatives at the edge of a tile, we load a set of neighboring points known as \textit{halo points}. We load the set of inner points and halo points from device memory to a 2D shared memory tile, evaluate the derivatives, and store the result back to shared memory. The inner points of one thread-block are halo-points of the adjacent thread-block and are loaded twice. We quantify this experimentally. We first repeat the FLOPS-MOPS experiment for the FD kernel and observe that the kernel is memory bound.

We compare the bandwidth performance of our general kernel to the parent SDK example. The SDK code works only for a fixed grid size $N=64^3$ and a 9-point stencil. CUDA SDK reports an effective bandwidth of 310\,GB/s whereas our implementation achieves 212\,GB/s. The reported bandwidth includes the cost of loading halo points. Both values are much smaller than $B_{\text{max}}$ because the grid size is not large enough to hide latency. Unlike the SDK example, the CUDA threads on the boundary of the domain load halo points from global memory instead of shared memory. The observed performance drops due to the thread divergence caused by reading out-of-bound halo points. For large $N$, as we show later, this overhead is greatly reduced as a direct consequence of decreased latency caused by higher occupancy.

We perform a zero-overhead memory copy, i.e., copy within the HBM2 device memory to put an absolute upper bound on the performance of our implementation. We load each element of an array of size $N=256^3$ from the global device memory and store it in another array. The peak performance we get for this copy routine is 780\,GB/s. To quantify the halo points load overhead, we perform another experiment. Each thread-block loads its inner points and halo points into a 2D shared memory tile and copies only the inner points back to the output array. The effective bandwidth for this benchmark is 766\,GB/s. The reported bandwidth includes the cost of loading halo points. We only lose 1.8\% of the memory bandwidth in comparison to the zero-overhead memory copy experiment. This indicates that the overhead due to loading of out-of-bound halo points gets smaller as the kernel occupancy increases. We verify our claims by profiling the kernels using the NVIDIA Visual Profiler. For the smaller grid size of $64^3$, the kernel is bound by instruction and memory latency, for the larger grids ($128^3$ and $256^3$) by memory bandwidth.

\begin{table}
\begin{minipage}[c]{0.53\textwidth}
\caption{Runtime (in seconds) of first order differential operators (gradient and divergence) using FFT and 8\textsuperscript{th} order finite differences (FD8) on a NVIDIA Tesla V100 for different problem sizes. We report the runtime in $s$ per kernel call averaged over the whole registration run from experiments shown in \secref{s:results} including all pre- and post-processing needed.
}
\end{minipage}
\begin{minipage}[c]{0.45\textwidth}
\centering\scriptsize\renewcommand{\arraystretch}{0.8}
\begin{tabular}{llcc}
\toprule
$N$ & {\bf Operator} & {\bf FFT} & {\bf FD8}\\
\midrule
\multirow{2}{*}{$64^3$}
& grad & \scnum{1.71E-04} &	\scnum{3.59E-05} \\
& div  & \scnum{1.68E-04} & \scnum{3.92E-05} \\ \midrule
\multirow{2}{*}{$128^3$}
& grad & \scnum{6.01E-04} & \scnum{1.36E-04} \\
& div  & \scnum{5.67E-04} & \scnum{1.59E-04} \\
\midrule
\multirow{2}{*}{$256^3$}
& grad & \scnum{4.05E-03} & \scnum{9.36E-04}  \\
& div  & \scnum{3.81E-03} & \scnum{1.16E-03} \\
\bottomrule
\end{tabular}
\end{minipage}
\end{table}

\section{IMAGE REGISTRATION RESULTS}\label{s:results}

We evaluate the overall algorithm using four 3D MRI images. We study convergence behavior, time-to-solution, and registration accuracy for several algorithmic variants of computational kernels available in our new GPU implementation of the CPU software \claire{}. We compare with two popular GPU packages for \lddmm{} registration. The purpose of this section is to show that (a) our new (mixed-precision) GPU implementation yields the same registration accuracy as our CPU implementation of \claire{}~\cite{Mang:2018CLAIRE} and (b) to compare our method against GPU implementations of other groups.

\subsection{Data and Setup}
\label{s:data_setup}

\subsubsection{Images}

We report results for the \sw{NIREP}  (Non-Rigid Image Registration Evaluation Project) data, a commonly used data set to evaluate the performance of deformable registration algorithms~\cite{Christensen:2006a}. NIREP consists of 16 rigidly aligned T1-weighted magnetic resonance neuroimaging MR scans (\texttt{na01}--\texttt{na16}) of different individuals. The original resolution is $256\times300\times256$ voxels. Each scan is annotated with a label map that identifies 32  gray matter regions~\cite{Christensen:2006a}. We select four scans from this data set,  \texttt{na01} as reference image and \texttt{na02}, \texttt{na03}, and \texttt{na10} as template images, respectively. The initial DICE coefficient (spatial overlap index) for the union of the gray matter regions of the template images versus the reference image is 0.55, 0.50 and 0.48, respectively. A perfect matching would correspond to a value of 1.00. Currently, we only support image sizes $N_1 N_2 N_3$ dividable by $256$. We resampled the data sets to grid sizes of $64^3$, $128^3$, $256^3$, and $384^3$, using a linear and a nearest-neighbor interpolation model for the image data and the label maps, respectively.

\begin{table}
\begin{minipage}[c]{0.45\textwidth}
\caption{Variants of combinations of computational kernels and the respective tag used in this work. IP stands for interpolation and FD8 for finite difference operators of 8\textsuperscript{th} order. \label{t:variants}}
\end{minipage}
\begin{minipage}[c]{0.5\textwidth}
\centering\scriptsize\renewcommand{\arraystretch}{0.8}
\begin{tabular}{ll} \toprule
Tag              & Variant                                \\\midrule
cpu-fft-cubic    & FP32, CPU, FFT, cubic IP \\
gpu-fft-cubic    & FP32, GPU, FFT, cubic IP              \\
gpu-fd8-cubic    & FP32, GPU, FD8, cubic IP              \\
gpu-fd8-linear   & FP32, GPU, FD8, trilinear IP          \\
\bottomrule
\end{tabular}
\end{minipage}
\end{table}

\subsubsection{Numerical \& Floating Point Accuracy Parameters}

Unless specified otherwise, we use the default solver parameters from~\cite{claire-web} for the Gauss--Newton--Krylov solver. For regularization we use the default of \claire{}, $H^1$-div---an $H^1$-seminorm with an additional penalty on the divergence of the velocity. In all runs, we use a target regularization parameter $\beta=\scnum{5e-4}$ selected based on experiments reported in~\cite{Mang:2018CLAIRE}. We execute the proposed solver with a parameter continuation scheme for the regularization parameter $\beta$. This scheme is describe in detail in~\cite{Mang:2015NK}. We set the parameter for the penalty for the divergence of $\vel$ to \scnum{1e-4}.

\begin{itemize}
\item {\bf Convergence Criteria:} As a stopping criterion for the optimizer, we use a tolerance of \scnum{5e-2} for the relative reduced gradient~\eqref{e:reducedgrad} together with a maximal number of Gauss--Newton iterations of $50$ (never reached in our experiments). We use a superlinear forcing sequence for the Newton-Krylov solver (inexact Newton solve; see \cite{Eisenstat:1996a,Dembo:1982a} for details) and set the maximum number of iterations for the PCG (used to compute the search direction; see \secref{s:methods}) to $500$ (never reached in our experiments). We globalize our Gauss--Newton--Krylov method using an Armijo line search~\cite{Nocedal:2006a}.
\item {\bf Interpolation:} We consider different interpolation methods to evaluate the value of variables at off grid locations within our semi-Lagrangian scheme (see \secref{s:methods}). In particular, we select either a linear or a cubic interpolation scheme. For cubic interpolation, we use GPU-TXTSPL as proposed in~\secref{s:kernels}.
\item {\bf First Order Derivatives:} For the calculation of first order derivatives, we compare the FFT-based scheme and the 8\textsuperscript{th} order finite difference (\acr{FD8}) scheme as proposed in~\secref{s:methods}.
\item {\bf Floating Point Accuracy:} Our new implementation uses single precision (\acr{FP32}). For validation, we compare against results achieved with the \claire{} CPU implementation in single precision. We summarize our settings in \tabref{t:variants}.
\end{itemize}

\subsubsection{Performance Metrics}

We report two groups of metrics: To assess computational performance, we report runtimes. To assess accuracy of the results, we report the relative mismatch $\|\sta(\,\bullet\,,1) - \sta_1\|_2 / \| \sta_1 - \sta_0 \|_2$ of the template image $\sta_0(\x)$, the reference image $\sta_1(\x)$, and the transformed template image $\sta(\x,1)$ given by the forward problem \eqref{e:varopt:constraint} as well as the DICE coefficient (overlap) between the union of the gray matter labels associated with the data sets. This enables an assessment of how well anatomical structures identified by expert observers are aligned after registration. For a perfect matching the value is 1.00\footnote{The DICE coefficient is a metric that has been widely adopted by the registration community to assess registration accuracy. We provide a more detailed study in \cite{Mang:2018CLAIRE}. We note that DICE and mismatch values do not provide a complete picture about registration accuracy. Other metrics include the Haussdorff distance between the contours of label maps or landmark errors (an example for a database that considers landmarks to evaluate registration performance is DIRLAB; see \texttt{www.dir-lab.com}). We note that the focus of the manuscript is on computational performance and not registration accuracy. The accuracy results included in this study serve as a baseline to compare our improved solver to our past work~ \cite{Mang:2018CLAIRE}.}. To measure the quality of the computed deformation map, we report min, mean and max values of the determinant of the deformation gradient $\detY$, $\mat{F}\in\ns{R}^{3,3}$.  The mapping is locally non-diffeomorphic if the determinant of the deformation gradient changes sign or is zero. In general, if $\detY$ is either very small (but still positive) or very big, the \lddmm{} mapping is of poor quality. In our case, $\detY$ is between $0.5$ and $10$, which indicates excellent registration quality.

To assess the (rate of) convergence of our solver, we report the relative gradient norm $\relG \defeq \|\vect{g}^\star\|_2 / \|\vect{g}^0\|_2$, where $\vect{g}^\star$ is the gradient of the optimization problem after convergence and $\vect{g}^0$ is the gradient for the initial guess $\vect{v} = \vect{0}$. We also report the number of iterations for the Newton-Krylov solver and the total number of Hessian matvecs (application of the Hessian to a vector; the smaller the better; see \secref{s:methods}).

\subsection{Results}
\label{s:reg_results}

Next, we report results for our improved implementation of \claire{}. We use the same experimental setup as for the kernel performance analysis in \secref{s:perf}.

\subsubsection{Performance Analysis of the Proposed Method}
\label{s:performance-reg}

\ipoint{Purpose:} We study the performance of different variants of our solver, i.e., for different combinations of computational kernels.

\begin{figure}
\centering
\resizebox{.9\linewidth}{!}{\input{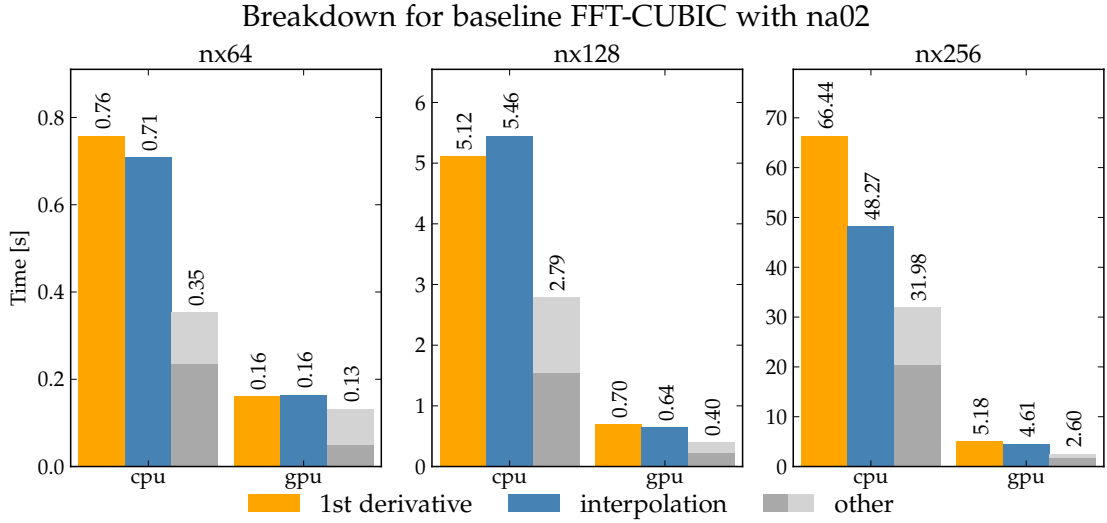}}
\caption{Runtime breakdown for the main kernels of the proposed method and the baseline CPU implementation in \claire{} (first order derivatives via FFT, cubic interpolation). The dark gray parts indicate the FFTs used for the regularization terms.
We consider the registration of the \texttt{na02} image to the \texttt{na01} image at a resolution of $64^3$, $128^3$, and $256^3$, respectively.
Note that the speed-up when moving to the GPU is a combination of algorithmic improvements and the higher memory bandwidth.}
\label{fig:breakdown-base}
\end{figure}

\begin{figure}
\centering
\resizebox{.9\linewidth}{!}{\input{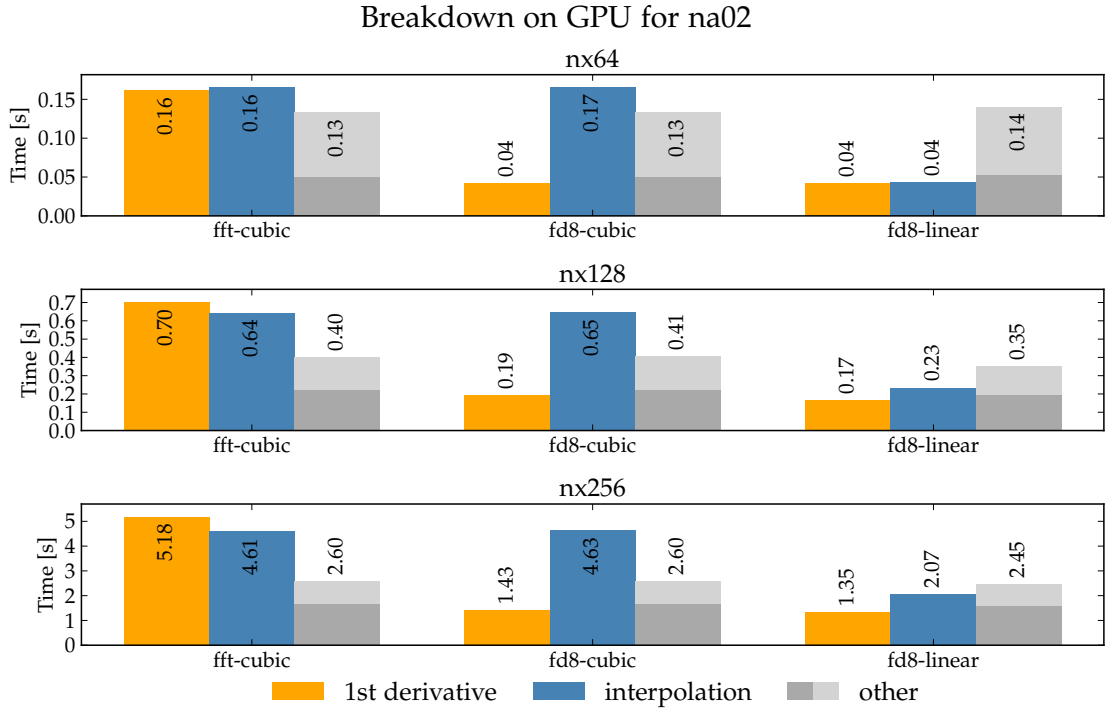}}
\caption{Runtime breakdown for the main kernels of the proposed method for all GPU implementations (first order derivatives via FFT or FD8, cubic or linear interpolation). The dark gray parts indicate the contribution of higher order operators in spectral space to the overall execution time of the solver.
We consider the registration of \texttt{na02} to \texttt{na01} at a resolution of $64^3$, $128^3$, and $256^3$, respectively.
}
\label{fig:breakdown}
\end{figure}

\ipoint{Results:} The results for the experiments described above are reported in \tabref{t:claire-runs} for image sizes of $64^3$, $128^3$, $256^3$, and $384^3$, respectively. The breakdown of the execution time with respect to the individual kernels is shown in \figref{fig:breakdown-base} and \figref{fig:breakdown}. \figref{fig:breakdown-base} compares runtimes between the  baseline CPU implementation with the equivalent GPU implementation using FFT for first order derivatives and cubic interpolation for the semi-Lagrangian scheme. We compare different GPU implementations in \figref{fig:breakdown} (for \texttt{na02}).
The maximum allocated memory on the GPU during the experiments was \SI{0.6}{\giga\byte}, \SI{1.3}{\giga\byte}, \SI{6.1}{\giga\byte}, and \SI{20.0}{\giga\byte} for image sizes of $64^3$, $128^3$, $256^3$, and $384^3$, respectively. The maximum allocated memory on the host CPU  was below \SI{2}{\giga\byte} for all GPU experiments and only used for management and IO purposes.

\ipoint{Observations:} The critical result is that we can accurately solve 3D image registration problems for clinically relevant sizes ($256^3$) on a single GPU in less than $10$ seconds (\runref{28}, \runref{32} and \runref{36} in \tabref{t:claire-runs}) for the variant gpu-fd8-linear. The gpu-fd8-cubic approximation is almost as fast while resulting in lower reduced gradient and  similarity than gpu-fd8-linear. We also found that the iteration counts, registration quality and number of Hessian matvecs remains almost constant as we switch to lower accuracy regimes. The values for the DICE, the relative mismatch between the deformed template image and the reference image, and the Gauss-Newton iteration counts are almost identical. We observe slight differences in the number of Hessian matvecs between implementations, with fewer matvecs typically observed for gpu-fd8-linear. For all implementations we reach the set tolerance of \scnum{5e-2} for the relative reduction of the gradient.

All implementations produce well-behaved determinants of the deformation gradients. The highest DICE score is achieved for \texttt{na02} (\num{0.86}, \runref{25} and \runref{28}). For gpu-fd8-linear, we see an increase in the maximum determinant of the deformation gradient, indicating a slightly more irregular mapping. For example, for \runref{12} or \runref{32} in \tabref{t:claire-runs}, the maximum of the determinant of the deformation gradient increases from $7.54$ to $10.52$ ($14$\%) and from $7.18$ to $7.92$ ($11$\%). The speedup between the baseline method cpu-fft-cubic and gpu-fd8-linear is $8$--$11$ for $64^3$, $16$--$18$ for $128^3$, and $23$--$25$ for $256^3$. The gpu-fd8-cubic variant also performs very well with similar run times and slightly better $\detY$.

For the considered test problems with image sizes $64^3$, $128^3$, and $256^3$, the number of Gauss--Newton iterations remains constant per resolution level with a minimum of $12$ and a maximum of $18$ Gauss--Newton iterations. The number of Hessian matvecs increases up to a factor of two as we change resolution levels, with a minimum of $42$ (\runref{8})  and a maximum of $104$ (\runref{34} and \runref{35}). There are several reasons for the increase in the number of matvecs. First, we can resolve finer details in the velocity and the images, which results in more complicated deformation patterns and by that longer runtimes. Second, we use a regularization parameter of $\beta=\scnum{1e-4}$ for all resolutions, to be consistent. Given the observed change of information content, one should in general adapt the regularization parameter according to the resolution level in real application cases. Our experiments for the image size $384^3$ have a higher variation in the number of Newton steps and matvecs. Notice that we use relative tolerances in our algorithm (as opposed to a fixed number of iterations). Consequently, we expect that differences in numerical accuracy and changes in the resolution (more frequencies can be resolved) have an effect on the number of iterations required until convergence.

\begin{table*}
\caption{Results for registration runs using the proposed method.
The experiments for the baseline (fft-cubic) implementation are highlighted in gray. We report for each dataset (from left to right): minimum, mean and maximum value of the determinant of the deformation gradient $\detY$, the DICE coefficient before and after registration, the relative mismatch, the relative $\ell^2$-norm of the gradient, the number of Gauss--Newton iterations until convergence (\#iter), the number of Hessian matvecs (\#MV), and the total runtime in seconds. We report results for data grid sizes of $64^3$, $128^3$, $256^3$, and $384^3$.\label{t:claire-runs}}
\resetrunid
\renewcommand{\arraystretch}{0.8}\centering\scriptsize
\begin{tabular}{lllrrrrrrrrrr}\toprule
& &&\multicolumn{3}{c}{$\detY$}&\multicolumn{2}{c}{\textbf{DICE}}\\
run & \textbf{variant}                   & \textbf{data} & \textbf{min} & \textbf{mean} & \textbf{max} & \textbf{before} & \textbf{after} & \textbf{mism.} & $\relG$ & \textbf{\#iter} & \textbf{\#MV} & \textbf{time} \\\midrule
 \rowcolor{colorA} \multicolumn{13}{l}{$N=64^3$}\\\midrule
 \rowcolor[gray]{0.8} \runid &  cpu-fft-cubic &  na02 & \num{0.643} & \num{1.008} &  \num{4.138} & \num{0.555} & \num{0.622} & \scnum{1.11e-02} & \scnum{7.74e-03} & \num{12} & \num{58} & \num{1.821}  \\
 \rowcolor[gray]{0.9} \runid &  gpu-fft-cubic &       & \num{0.634} & \num{1.008} &  \num{3.985} &             & \num{0.620} & \scnum{1.10e-02} & \scnum{9.03e-03} & \num{12} & \num{58} & \num{0.459}  \\
                      \runid &  gpu-fd8-cubic &       & \num{0.634} & \num{1.008} &  \num{3.963} &             & \num{0.620} & \scnum{1.10e-02} & \scnum{8.90e-03} & \num{12} & \num{58} & \num{0.339}  \\
                      \runid & gpu-fd8-linear &       & \num{0.643} & \num{1.009} &  \num{5.060} &             & \num{0.628} & \scnum{1.72e-02} & \scnum{1.08e-02} & \num{12} & \num{54} & \num{0.225}  \\
 \midrule
 \rowcolor[gray]{0.8} \runid &  cpu-fft-cubic &  na03 & \num{0.631} & \num{1.014} &  \num{8.497} & \num{0.503} & \num{0.614} & \scnum{8.67e-03} & \scnum{7.95e-03} & \num{13} & \num{64} & \num{1.970}  \\
 \rowcolor[gray]{0.9} \runid &  gpu-fft-cubic &       & \num{0.628} & \num{1.015} &  \num{8.036} &             & \num{0.612} & \scnum{8.62e-03} & \scnum{8.26e-03} & \num{13} & \num{63} & \num{0.539}  \\
                      \runid &  gpu-fd8-cubic &       & \num{0.629} & \num{1.015} &  \num{8.013} &             & \num{0.612} & \scnum{8.62e-03} & \scnum{8.20e-03} & \num{13} & \num{63} & \num{0.388}  \\
                      \runid & gpu-fd8-linear &       & \num{0.591} & \num{1.018} &  \num{9.061} &             & \num{0.613} & \scnum{1.41e-02} & \scnum{1.60e-02} & \num{12} & \num{42} & \num{0.177}  \\
 \midrule
 \rowcolor[gray]{0.8}  \runid & cpu-fft-cubic &  na10 & \num{0.558} & \num{1.032} &  \num{7.881} & \num{0.479} & \num{0.678} & \scnum{7.16e-03} & \scnum{1.23e-02} & \num{12} & \num{48} & \num{1.610} \\
 \rowcolor[gray]{0.9}  \runid & gpu-fft-cubic &       & \num{0.562} & \num{1.032} &  \num{7.475} &             & \num{0.677} & \scnum{7.06e-03} & \scnum{1.26e-02} & \num{12} & \num{48} & \num{0.408} \\
                       \runid & gpu-fd8-cubic &       & \num{0.561} & \num{1.032} &  \num{7.537} &             & \num{0.677} & \scnum{7.05e-03} & \scnum{1.25e-02} & \num{12} & \num{48} & \num{0.308} \\
                       \runid &gpu-fd8-linear &       & \num{0.587} & \num{1.031} & \num{10.516} &             & \num{0.684} & \scnum{9.59e-03} & \scnum{1.31e-02} & \num{12} & \num{44} & \num{0.176} \\
 \midrule
 \rowcolor{colorA} \multicolumn{13}{l}{$N=128^3$}\\
 \midrule
 \rowcolor[gray]{0.8}  \runid & cpu-fft-cubic &  na02 & \num{0.539} & \num{1.010} &  \num{3.986} & \num{0.553} & \num{0.793} & \scnum{1.70e-02} & \scnum{1.77e-02} & \num{14} & \num{70} & \num{13.364} \\
 \rowcolor[gray]{0.9}  \runid & gpu-fft-cubic &       & \num{0.539} & \num{1.014} &  \num{3.916} &             & \num{0.792} & \scnum{1.72e-02} & \scnum{1.78e-02} & \num{14} & \num{73} & \num{ 1.747} \\
                       \runid & gpu-fd8-cubic &       & \num{0.540} & \num{1.014} &  \num{3.924} &             & \num{0.792} & \scnum{1.72e-02} & \scnum{1.76e-02} & \num{14} & \num{73} & \num{ 1.245} \\
                       \runid &gpu-fd8-linear &       & \num{0.575} & \num{1.011} &  \num{4.793} &             & \num{0.797} & \scnum{2.02e-02} & \scnum{1.71e-02} & \num{12} & \num{63} & \num{ 0.751} \\
 \midrule
 \rowcolor[gray]{0.8}  \runid & cpu-fft-cubic &  na03 & \num{0.478} & \num{1.017} &  \num{8.103} & \num{0.505} & \num{0.785} & \scnum{1.53e-02} & \scnum{1.82e-02} & \num{15} & \num{77} & \num{14.619} \\
 \rowcolor[gray]{0.9}  \runid & gpu-fft-cubic &       & \num{0.483} & \num{1.024} &  \num{7.932} &             & \num{0.785} & \scnum{1.57e-02} & \scnum{1.89e-02} & \num{15} & \num{78} & \num{ 1.855} \\
                       \runid & gpu-fd8-cubic &       & \num{0.484} & \num{1.024} &  \num{7.935} &             & \num{0.785} & \scnum{1.57e-02} & \scnum{1.87e-02} & \num{15} & \num{78} & \num{ 1.332} \\
                       \runid &gpu-fd8-linear &       & \num{0.483} & \num{1.021} & \num{10.143} &             & \num{0.790} & \scnum{1.58e-02} & \scnum{1.74e-02} & \num{13} & \num{68} & \num{ 0.813} \\
 \midrule
 \rowcolor[gray]{0.8}  \runid & cpu-fft-cubic &  na10 & \num{0.545} & \num{1.036} &  \num{8.779} & \num{0.479} & \num{0.776} & \scnum{1.19e-02} & \scnum{1.65e-02} & \num{15} & \num{84} & \num{15.931} \\
 \rowcolor[gray]{0.9}  \runid & gpu-fft-cubic &       & \num{0.572} & \num{1.036} &  \num{8.859} &             & \num{0.775} & \scnum{1.18e-02} & \scnum{1.62e-02} & \num{14} & \num{82} & \num{ 1.921} \\
                      \runid & gpu-fd8-cubic &       & \num{0.572} & \num{1.036} &  \num{8.844} &             & \num{0.775} & \scnum{1.18e-02} & \scnum{1.61e-02} & \num{14} & \num{82} & \num{ 1.358} \\
                      \runid &gpu-fd8-linear &       & \num{0.579} & \num{1.031} &  \num{9.981} &             & \num{0.780} & \scnum{1.30e-02} & \scnum{1.66e-02} & \num{15} & \num{82} & \num{ 0.963} \\
 \midrule
 \rowcolor{colorA} \multicolumn{13}{l}{$N=256^3$}\\
 \midrule
 \rowcolor[gray]{0.8}  \runid &  cpu-fft-cubic &  na02 & \num{0.411} & \num{1.012} & \num{3.616} & \num{0.554} & \num{0.855} & \scnum{2.89e-02} & \scnum{3.67e-02} & \num{14} & \num{81} & \num{146.685} \\
 \rowcolor[gray]{0.9}  \runid &  gpu-fft-cubic &       & \num{0.413} & \num{1.005} & \num{3.570} &             & \num{0.854} & \scnum{2.98e-02} & \scnum{3.81e-02} & \num{14} & \num{81} & \num{ 12.383} \\
                       \runid &  gpu-fd8-cubic &       & \num{0.414} & \num{1.005} & \num{3.567} &             & \num{0.854} & \scnum{2.98e-02} & \scnum{3.74e-02} & \num{14} & \num{81} & \num{  8.657} \\
                       \runid & gpu-fd8-linear &       & \num{0.426} & \num{1.009} & \num{3.829} &             & \num{0.858} & \scnum{2.73e-02} & \scnum{3.09e-02} & \num{14} & \num{75} & \num{  5.874} \\
 \midrule
 \rowcolor[gray]{0.8}  \runid &  cpu-fft-cubic &  na03 & \num{0.474} & \num{1.019} & \num{6.832} & \num{0.504} & \num{0.828} & \scnum{2.81e-02} & \scnum{3.63e-02} & \num{17} & \num{95} & \num{169.457} \\
 \rowcolor[gray]{0.9}  \runid &  gpu-fft-cubic &       & \num{0.468} & \num{1.001} & \num{6.809} &             & \num{0.827} & \scnum{2.88e-02} & \scnum{3.77e-02} & \num{17} & \num{99} & \num{ 15.093} \\
                       \runid &  gpu-fd8-cubic &       & \num{0.468} & \num{1.001} & \num{6.789} &             & \num{0.827} & \scnum{2.88e-02} & \scnum{3.69e-02} & \num{17} & \num{98} & \num{ 10.440} \\
                       \runid & gpu-fd8-linear &       & \num{0.478} & \num{1.001} & \num{7.511} &             & \num{0.832} & \scnum{2.55e-02} & \scnum{3.11e-02} & \num{17} & \num{93} & \num{  7.224} \\
 \midrule
 \rowcolor[gray]{0.8}  \runid &  cpu-fft-cubic &  na10 & \num{0.579} & \num{1.036} & \num{7.183} & \num{0.479} & \num{0.816} & \scnum{2.07e-02} & \scnum{3.52e-02} & \num{18} & \num{103} & \num{184.776} \\
 \rowcolor[gray]{0.9}  \runid &  gpu-fft-cubic &       & \num{0.578} & \num{1.011} & \num{7.084} &             & \num{0.815} & \scnum{2.16e-02} & \scnum{3.79e-02} & \num{18} & \num{104} & \num{ 16.054} \\
                       \runid &  gpu-fd8-cubic &       & \num{0.577} & \num{1.011} & \num{7.183} &             & \num{0.816} & \scnum{2.08e-02} & \scnum{3.36e-02} & \num{18} & \num{104} & \num{ 11.046} \\
                       \runid & gpu-fd8-linear &       & \num{0.605} & \num{1.010} & \num{7.917} &             & \num{0.818} & \scnum{1.96e-02} & \scnum{2.94e-02} & \num{17} & \num{ 94} & \num{  7.292} \\ \midrule
\rowcolor{colorA} \multicolumn{13}{l}{$N=384^3$}\\ \midrule
\rowcolor[gray]{0.9}  \runid & gpu-fft-cubic  &  na02 & \num{0.366} & \num{0.593} & \num{3.780} & \num{0.554} & \num{0.861} & \scnum{2.61e-02} & \scnum{3.35e-02} & \num{16} & \num{152} & \num{72.820} \\
                      \runid & gpu-fd8-cubic  &       & \num{0.402} & \num{0.593} & \num{3.551} &             & \num{0.851} & \scnum{3.36e-02} & \scnum{4.34e-02} & \num{15} & \num{91} & \num{31.590} \\
                      \runid & gpu-fd8-linear &       & \num{0.410} & \num{0.593} & \num{3.707} &             & \num{0.854} & \scnum{3.06e-02} & \scnum{3.75e-02} & \num{15} & \num{85} & \num{21.692} \\ \midrule
\rowcolor[gray]{0.9} \runid &  gpu-fft-cubic  &  na03 & \num{0.458} & \num{0.599} & \num{7.523} & \num{0.504} & \num{0.835} & \scnum{2.71e-02} & \scnum{4.27e-02} & \num{22} & \num{201} & \num{96.594} \\
                     \runid & gpu-fd8-cubic  &        & \num{0.437} & \num{0.600} & \num{6.630} &             & \num{0.826} & \scnum{3.25e-02} & \scnum{4.12e-02} & \num{18} & \num{112} & \num{38.719} \\
                     \runid & gpu-fd8-linear &        & \num{0.451} & \num{0.599} & \num{6.989} &             & \num{0.827} & \scnum{3.00e-02} & \scnum{3.78e-02} & \num{17} &  \num{98} & \num{24.901} \\  \midrule
\rowcolor[gray]{0.9} \runid &  gpu-fft-cubic &  na10  & \num{0.585} & \num{0.608} & \num{7.976} & \num{0.479} & \num{0.812} & \scnum{2.19e-02} & \scnum{3.79e-02} & \num{25} & \num{233} & \num{111.545} \\
                     \runid &  gpu-fd8-cubic &        & \num{0.552} & \num{0.609} & \num{7.201} &             & \num{0.804} & \scnum{2.58e-02} & \scnum{4.20e-02} & \num{20} & \num{117} & \num{40.816} \\
                     \runid &  gpu-fd8-linear &       & \num{0.575} & \num{0.606} & \num{7.492} &             & \num{0.805} & \scnum{2.39e-02} & \scnum{3.68e-02} & \num{18} & \num{104} & \num{26.351} \\
\bottomrule
\end{tabular}
\end{table*}

Looking at the breakdown of the CPU baseline in \figref{fig:breakdown-base}, we observe that its runtime is dominated by the application of first-order derivatives and interpolation operations. If we add the execution time of high-order spectral derivatives (bars in dark gray in the "other" category), we see that almost all runtime goes to differentiation and interpolation. We spend $\SI{66.44}{\second} + \SI{48.27}{\second} = \SI{114.71}{\second}$ out of \SI{146.69}{\second} ($78$\% of the runtime) on computing first-order derivatives and evaluating the interpolation kernel (right plot in \figref{fig:breakdown-base}; CPU; grid size: $256^3$). We observe a similar behavior for the GPU implementation. For example, we spend $\SI{5.18}{\second}+\SI{4.61} = \SI{9.79}{\second}$ of \SI{12.39}{\second} ($80$\% of the runtime)  on these kernels (right plot in \figref{fig:breakdown-base}; GPU; grid size: $256^3$). Consequently, we expect a significant reduction in the runtime of our GPU accelerated version of \claire{} compared to the CPU implementation of \claire{} if we can speed up the evaluation of these kernels. This is precisely what we observe in \tabref{t:claire-runs}.

The breakdown in \figref{fig:breakdown} provides additional insight. We can see that the execution time for the first-order derivatives reduces from \SI{5.18}{\second} to \SI{1.43}{\second} (speed up of $\approx 3.5$) when switching from spectral methods to an optimized FD8 approximation (\figref{fig:breakdown}, bottom block; yellow bars for the 1st derivative). If we switch from cubic to linear interpolation, we see a reduction in the execution time from \SI{4.63}{\second} to \SI{2.07}{\second} (speed up of $\approx 2$). The runtime of the other operations remains almost constant. So, overall we went from a solver that is bound by the through-put of first order derivatives and interpolation operations, to a solver that is now bound by the execution time of high-order derivatives.

\begin{table}
\caption{Registration performance for \sw{PyCA} \cite{pyca-git}, \sw{deformetrica} \cite{deformetrica-git}, and the proposed method executed on a V100 and a P100 for three neuroimaging data sets (grid size: $256^3$). We were not able to execute \sw{deformetrica} on a V100 due to issues with the installation. We expect the speedup to be 2$\times$ (in accordance with the observations we have made for the other software packages); \sw{deformetrica} would still be slower than \sw{PyCA}. The solvers are executed with default parameters. We only alter the maximum number of iterations. The defaults are 300 iterations per level for \sw{PyCA} (using a multi-resolution strategy with two levels) and 50 iterations for \sw{deformetrica}. We execute the proposed method with a parameter continuation scheme for the regularization parameter (the default method used in the CPU version of \claire{}); we report results for the proposed method corresponding to \runref{28}, \runref{32}, and \runref{36}, in \tabref{t:claire-runs}. We report iterations per level (``100,50'' for PyCA means 100 iterations on the first level and 50 iterations on the second level), the relative mismatch after registration (mism.), and the runtime (in seconds). We see that {\bf our GPU implementation of \claire{} is about an order of magnitude more accurate (mismatch) and, at the same time, up to 30$\times$ faster} (fastest result for \sw{PyCA} on a V100). The runs \#3/14/19 for CLAIRE correspond to the runs \#28/32/36 in \tabref{t:claire-runs} (same experiment). \label{t:competitors}}
\resetrunid\scriptsize\centering
\renewcommand{\arraystretch}{0.8}
\begin{tabular}{r|rrrrr|rrrrr|rrrrrr}\toprule
      & \multicolumn{5}{r|}{\sw{PyCA} \cite{pyca-git}}                                                                         & \multicolumn{5}{r|}{\sw{deformetrica} \cite{deformetrica-git}}             & \multicolumn{5}{r}{proposed method}                                            \\\midrule
data  & run    & \textbf{\#iter}       & \textbf{mism.}       & \textbf{time}             &                           & run    & \textbf{\#iter} & \textbf{mism.}       & \textbf{time}          &          & run     & \textbf{\#iter} & \textbf{mism.}    & \textbf{time}  &               \\\midrule
      &        &                       &                      & P100                      &                     V100  &        &                 &                      & P100                   & V100     &         &                 &                   & P100           & V100          \\\midrule
na02  & \runid & \num{100},\num{50}    & \scnum{4.176573e-01} & \scnum{1.89144029617e+01} & \scnum{1.08566408157e+01} & \runid & \num{ 10}       & \scnum{4.803172e-01} & \scnum{1.36e+02}       & --       & \runid  & \num{14}        & \scnum{2.70e-02}  & \scnum{9.0135} & \scnum{5.874} \\
      & \runid & \num{100},\num{100}   & \scnum{3.435998e-01} & \scnum{3.35074269772e+01} & \scnum{1.78387768269e+01} & \runid & \num{ 25}       & \scnum{3.970968e-01} & \scnum{2.48e+02}       & --       &         &                 &                                                    \\
      & \runid & \num{300},\num{300}   & \scnum{2.401902e-01} & \scnum{1.00769843102e+02} & \scnum{5.30431728363e+01} & \runid & \num{ 50}       & \scnum{3.456086e-01} & \scnum{4.36e+02}       & --       &         &                 &                                                    \\
      & \runid & \num{500},\num{500}   & \scnum{2.055471e-01} & \scnum{1.68958894968e+02} & \scnum{8.86050348282e+01} & \runid & \num{100}       & \scnum{3.161754e-01} & \scnum{8.21e+02}       & --       &         &                 &                                                    \\
      & \runid & \num{1000},\num{1000} & \scnum{1.856100e-01} & \scnum{3.41336262941e+02} & \scnum{1.78770730972e+02} & \runid & \num{300}       & \scnum{2.774859e-01} & \scnum{2.35e+03}       & --       &         &                 &                                                    \\\midrule
na03  & \runid & \num{300},\num{300}   & \scnum{2.489134e-01} & \scnum{1.00708877802e+02} & \scnum{5.38157811165e+01} & \runid & \num{ 50}       & \scnum{3.095330e-01} & \scnum{8.38e+02}       & --       & \runid  & \num{17}        & \scnum{2.55e-02}  & \scnum{11.0857} & \num{7.224}  \\
      & \runid & \num{500},\num{500}   & \scnum{2.468163e-01} & \scnum{1.69317574024e+02} & \scnum{9.00696249008e+01} & \runid & \num{300}       & \scnum{2.484198e-01} & \scnum{2.38e+03}       & --       &         &                 &                                                    \\\midrule
na10  & \runid & \num{300},\num{300}   & \scnum{2.476202e-01} & \scnum{1.00524739027e+02} & \scnum{5.38444380760e+01} & \runid & \num{ 50}       & \scnum{2.985191e-01} & \scnum{8.28e+02}       & --       & \runid  & \num{17}        & \scnum{1.96e-02}  & \scnum{11.172}  & \num{7.292}  \\
      & \runid & \num{500},\num{500}   & \scnum{2.198946e-01} & \scnum{1.68389488935e+02} & \scnum{8.99167110920e+01} & \runid & \num{300}       & \scnum{2.520326e-01} & \scnum{2.39e+03}       & --       &         &                 &                                                    \\
\bottomrule
\end{tabular}
\end{table}

\begin{figure}
\centering
\includegraphics[width=\textwidth]{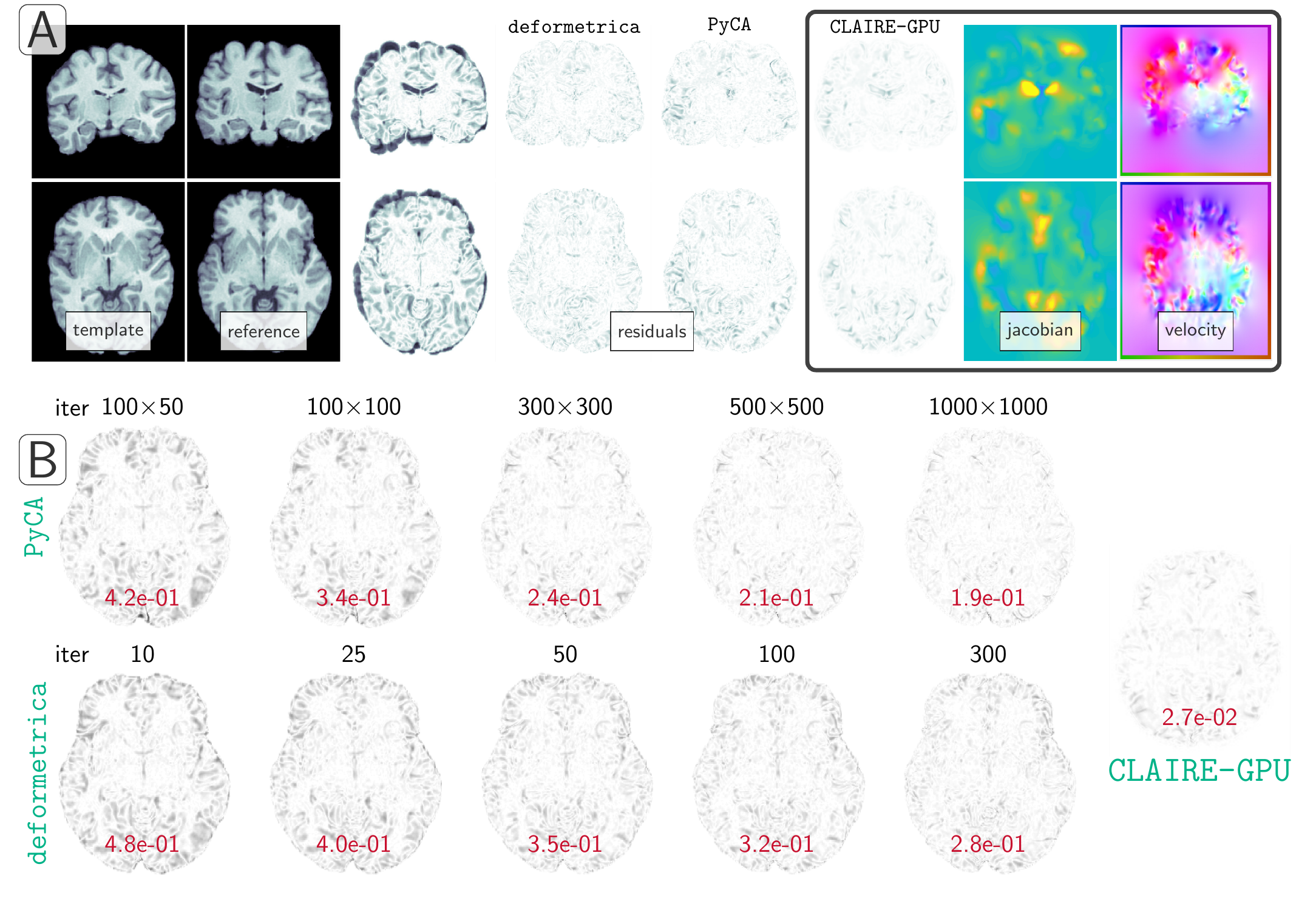}
\caption{Registration results. (A) We visualize the registration results for image \texttt{na03} to \texttt{na01}. Top row: Coronal view. Bottom row: Axial view. We show (from left to right) the template image $\sta_0(\x)$, the reference image $\sta_1(\x)$, the mismatch before registration, the mismatch after registration (for \sw{deformetrica}, \sw{PyCA}, and our improved implementation of \claire{}, respectively), and the determinant of the deformation gradient as well as the scalar map for the  orientation of the computed velocity vectors. The color bar for the values for the determinant of the deformation gradient is limited to $[0,2]$ with blue/green/yellow corresponding to 0\,/\,$\approx$1\,/\,$\geq$2 (values $\geq2$ are set to 2 for visualization purposes). The computed deformation map is locally diffeomorphic as judged by the determinant of the deformation gradient (up to numerical accuracy; min: \scnum{4.777575e-01}; max: \scnum{7.510764e+00}; mean: \scnum{1.018400e+00}). The results reported in this figure are the best-performing runs of those reported in \tabref{t:competitors} for each software. (B) Registration results for the image  \texttt{na02} to \texttt{na01}. We show results for different iteration settings for \sw{PyCA} (top row) and \sw{deformetrica} (bottom row). Results for \claire{} are shown on the right. The numbers in red are the obtained mismatch values for the respective settings.\label{f:claire-versus-the-world}}
\end{figure}

\begin{figure}
\centering
\includegraphics[width=\textwidth]{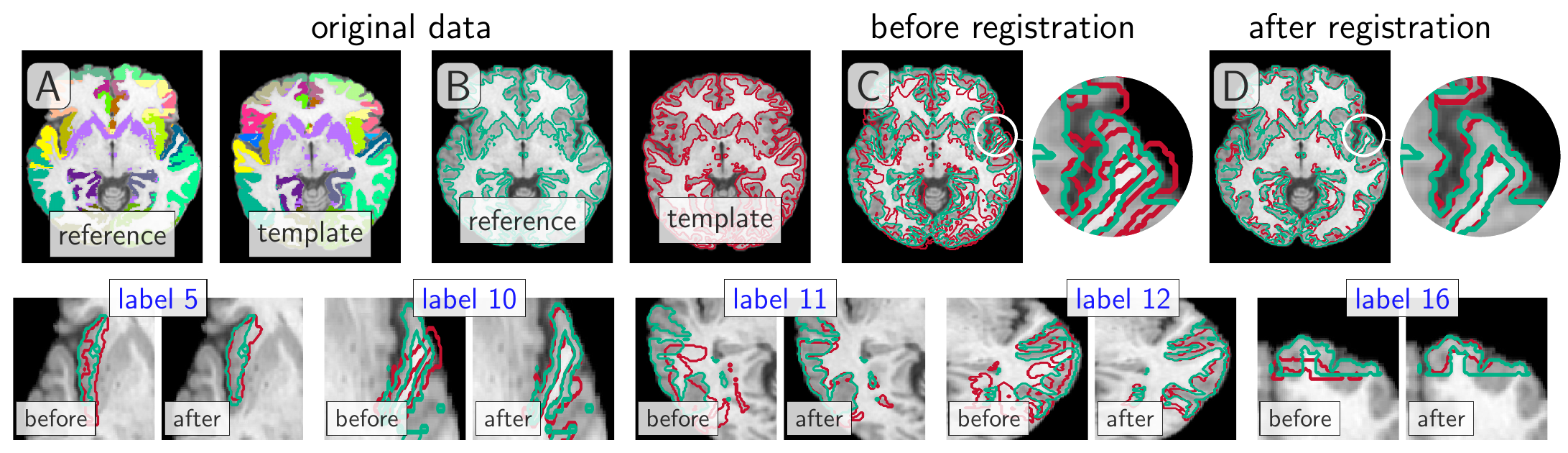}
\caption{Registration results for CLAIRE. Top row: In (A) we show the image data overlaid with the 32 gray matter labels (datasets \texttt{na03} and \texttt{na01}). In (B) we show the contours of the union of these labels overlaid onto the reference and template image, respectively. In (C) we show the two contours overlaid onto the reference image before registration and in (D) after registration (red contour: template image; green contour: reference image). The circles show a closeup. In the bottom row we show contours before and after registration (left and right, respectively) for five of the 32 gray matter labels visualized in (A) (top row).\label{f:claire-contour-plots}}
\end{figure}

\subsubsection{Comparison with other GPU Implementations}
\label{s:other}

\ipoint{Purpose:} We compare the performance of our new, improved GPU version of \claire{} to other GPU implementations of \lddmm{}-type methods.

\ipoint{Setup:} We compare the performance of the proposed method to publicly available GPU implementations of \lddmm{} approaches that have recently been considered by several groups~\cite{Bone:2018a,Bone:2018b,Fishbaugh:2017a,Yang:2016a,Yang:2017a}. The first software package is~\sw{PyCA}~\cite{pyca-git}. \sw{PyCA} uses gradient descent for optimization. Its interface is written in \sw{python}. The libraries and modules used for the compilation of \sw{PyCA} and \sw{deformetrica} are listed in the citations~\cite{pyca-git} and~\cite{deformetrica-git}, respectively. The second software package is \sw{deformetrica}~\cite{deformetrica-git}; \sw{deformetrica} uses a limited\--memory Broy\-den\--Fletcher\--Gold\-farb\--Shanno method for optimization. The gradient of the optimization problem is computed based on automatic differentiation~\cite{Bone:2018b}. We execute both registration packages for the three neuroimaging data sets we used to assess the performance of the proposed method (\texttt{na02}, \texttt{na03}, and \texttt{na10} as template images and \texttt{na02} as reference image). The runs are performed using the full resolution of our data ($256^3$). We slightly modify scripts available in the repositories of these two software packages to execute these runs (using the default parameters available in the scripts). We vary the number of iterations for \sw{PyCA} and \sw{deformetrica} to make sure we (i) do not terminate early, (ii) do not perform unnecessary iterations, and (iii) (possibly) generate the most accurate results attainable for the default settings (subject to a reasonable iteration count/runtime). We compare these results to our fastest implementation of the proposed method (gpu-fd8-linear; see results reported in \tabref{t:claire-runs}).

\ipoint{Results:} In~\tabref{t:competitors}, we report runtimes and relative mismatch values for all methods. We compare these results to the best performance achieved for the proposed method for the experiments reported in \tabref{t:claire-runs} (\runref{28}, \runref{32}, and \runref{36}). We showcase exemplary registration results as well as the imaging data to be registered in \figref{f:claire-versus-the-world} and \figref{f:claire-contour-plots}. In \figref{f:claire-versus-the-world}, we show (from left to right; coronal views: top row; axial views: bottom row) the reference image, the template image, the initial mismatch before registration, and the mismatch after registration for \sw{deformetrica}, \sw{PyCA}, and the proposed method, respectively. We also provide point wise maps for the determinant of the deformation gradient and a map of the orientation of the velocity field for the proposed method. \figref{f:claire-contour-plots} shows image data overlaid with the 32 gray matter labels, contours of the union of these labels overlaid onto the reference and template image, respectively, and overlaid contours before and after registration. We have reported extensive experiments in our past work~\cite{Mang:2018CLAIRE}. In the present work, we are only interested in demonstrating that switching to our GPU implementation (with mixed-precision accuracy) does not deteriorate the results we get.

\ipoint{Observations:} The most important observation is that the proposed method delivers a mismatch that is about one order of magnitude better than \sw{PyCA} and \sw{deformetrica} for the default settings, with more than one order of magnitude decrease in runtime. For the peak performance of the proposed method, we see that our approach is 30$\times$ faster with a 6$\times$ better mismatch (comparison of \runref{9} in \tabref{t:competitors} with the best result obtained for the proposed method; \runref{28} in \tabref{t:claire-runs}). Note that PyCA uses first order methods for optimization. Therefore, each iteration is much cheaper. In CLAIRE, we use second order information (Newton). Our method makes more progress per iteration but also requires more work; we need to iteratively invert the Hessian matrix to compute the search direction (i.e., solve a linear system). Thus, time per iteration is not a good measure on its own. We need to compare how much work (runtime) it requires to reach a certain accuracy (mismatch between the data). For the proposed method, we use convergence criteria based on the relative reduction of the gradient norm. The two other methods considered here terminate when they reach the set upper bound for the iterations. The best result is obtained for \sw{PyCA} with 1,000 gradient descent steps per level. If we would further increase the runtime (number of iterations) we would probably obtain results that are closer to those obtained for the proposed method (in terms of mismatch). We observe a linear increase in the runtime with respect to the number of iterations for both considered methods. We note that the differences in accuracy between the methods can be attributed to various factors (e.g., different optimization methods; convergence criteria; different regularization weights and norms; different parameters for the algorithm; or different mathematical formulations). The findings reported here are in accordance with timings reported in the literature~\cite{Bone:2018a,Yang:2016a,Yang:2017a}. \figref{f:claire-contour-plots} shows that not only the DICE coefficients indicate good quality od registration results, but also the label contours match very well after registration.

\section{CONCLUSIONS} \label{s:conclusions}

We presented algorithms, analysis, and numerical experiments for an improved GPU implementation of the CPU registration package \claire{} for large deformation diffeomorphic image registration. This problem is resource constrained because clinical workflows require high-throughput, with one or more registration tasks per node. Typical image sizes fit into the memory of a single GPU in our optimized implementation. MPI parallelism cannot help since multiple registration tasks can take place in an embarrassingly parallel way. Therefore, our focus is on single node and, in particular, on single device optimizations. We demonstrated over 10$\times$ speedup over state-of-the-art GPU implementations of \lddmm{} registration. We showed that the problem is memory-bound but it utilizes over 50\% of the peak bandwidth and has sufficient arithmetic intensity to deliver multi TFLOP/s performance.



\end{document}

%% file: preamble.tex
\usepackage[T1]{fontenc}

\usepackage[sort,compress]{cite}
\usepackage[ruled,noend,linesnumbered,algo2e]{algorithm2e}
\usepackage{amsmath,amssymb,amsfonts,amsbsy,amsfonts,latexsym}
\usepackage{multirow}
\usepackage{makecell}
\usepackage{subcaption}
\usepackage[labelfont=bf,textfont=it,belowskip=0pt,aboveskip=5pt,tableposition=top]{caption}
\usepackage{colortbl}
\usepackage{siunitx}
\usepackage{booktabs}
\usepackage[sc]{mathpazo}
\usepackage{graphicx}
\usepackage{rotating}
\usepackage{paralist}
\usepackage{lscape}
\usepackage{nicefrac}
\usepackage{geometry}
\usepackage[hyperpageref]{backref}
\usepackage[table]{xcolor}
\usepackage{tikz,pgfplots}
\usepackage{enumitem}
\usepackage{xr}

\newcounter{runidnum}
\definecolor{colorA}{RGB}{189,201,225}
\definecolor{colorB}{RGB}{103,169,207}
\definecolor{colorC}{RGB}{ 28,144,153}
\definecolor{colorD}{RGB}{  1,108, 89}

\newcolumntype{R}{>{\columncolor{gray!40}}r}
\newcolumntype{L}{>{\columncolor{gray!40}}l}
\newcolumntype{C}{>{\columncolor{gray!40}}c}

\newcommand{\runid}{\stepcounter{runidnum}\#\num{\therunidnum}}
\newcommand{\resetrunid}{\setcounter{runidnum}{0}}

\graphicspath{{./figs/}}
\DeclareGraphicsExtensions{.pdf,.png}

\usepackage{pgfplots, pgfplotstable}

\let\pgfimageWithoutPath\pgfimage
\renewcommand{\pgfimage}[2][]{\pgfimageWithoutPath[#1]{figs/#2}}

\DeclareCaptionLabelFormat{andtable}{#1~#2  \&  \tablename~\thetable}

\newcommand{\bvec}[1]{\mathbf{#1}}

\def\algclr#1{\leavevmode\rlap{\hbox to \hsize{\color{#1!30}\leaders\hrule height .8\baselineskip depth .5ex\hfill}}}
\def\algcomment#1{{\hfill$\triangleright$\small\textsf{ #1 }}}

%% file: macros.tex
\usepackage{xparse}
\NewDocumentCommand{\var}{O{s} m O{}}{%
  \ensuremath{#1_{#2}^{#3}}
}

\newcommand\scnum[1]{\num[scientific-notation=true,round-precision=1]{#1}}

\newcommand{\figref}[1]{Figure~\ref{#1}}
\newcommand{\tabref}[1]{Table~\ref{#1}}
\newcommand{\secref}[1]{\S\ref{#1}}

\newcommand{\runref}[1]{Run~\##1}

\newcommand{\vect}[1]{\boldsymbol{#1}} 
\newcommand{\mat}[1]{\boldsymbol{#1}}  
\newcommand{\defeq}{\ensuremath{\mathrel{\mathop:}=}}

\newcommand{\iquote}[1]{`'\textit{#1}`'}
\newcommand{\acr}[1]{\textbf{#1}}

\newcommand{\sw}[1]{\texttt{#1}}
\newcommand{\claire}{\texttt{CLAIRE}}
\newcommand{\lddmm}{LDDMM}

\DeclareMathOperator*{\minopt}{minimize}

\DeclareMathOperator{\bigO}{\mathcal{O}}

\newcommand{\ipoint}[1]{\textit{\textbf{\color{darkgray}#1}}}

\newcommand{\ns}[1]{\ensuremath{\mathbb{#1}}}
\newcommand{\fun}[1]{\ensuremath{\mathcal{#1}}}
\newcommand{\idiv}{\ensuremath{\nabla\cdot}}
\newcommand{\igrad}{\ensuremath{\nabla}}

\newcommand{\ts}{\textsuperscript}
\newcommand{\half}[1]{\frac{#1}{2}}

\newcommand{\dop}[1]{\ensuremath{\mathcal{#1}}}
\newcommand{\di}[1]{\ensuremath{\mathbf{#1}}}

\newcommand\x{\ensuremath{\vect{x}}}
\newcommand\sta{\ensuremath{m}}
\newcommand\vel{\ensuremath{\vect{v}}}
\newcommand\dmap{\ensuremath{\vect{y}}}
\newcommand\adj{\ensuremath{\lambda}}

\newcommand\traj{\ensuremath{\vect{y}}}

\newcommand\divel{\ensuremath{\di{\tilde{v}}}}
\newcommand\dvel{\ensuremath{\di{v}}}

\renewcommand{\d}[1]{\mathop{}\!\mathrm{d}#1}
\newcommand{\dt}{\d{t}}
\newcommand{\dx}{\d{\x}}
\newcommand{\p}{\partial}

\newcommand{\relG}{\ensuremath{\|\vect{g}\|_{\text{rel}}}}
\newcommand{\detY}{\ensuremath{\det\mat{F}}}

\newcommand{\bipa}{\begin{inparaenum}[(\itshape i\upshape)]}
\newcommand{\eipa}{\end{inparaenum}}
\newcommand{\bipasub}{\begin{inparaenum}[(\itshape a\upshape)]}
\newcommand{\eipasub}{\end{inparaenum}}

\definecolor{light-gray}{gray}{0.80}

\renewcommand\paragraph{\subsubsection*}

%% file: paper.bbl
\begin{thebibliography}{10}

\bibitem{cuda-docs}
{\em Cuda toolkit dcoumentation}.

\bibitem{Arsigny:2006a}
{\sc V.~Arsigny, O.~Commowick, X.~Pennec, and N.~Ayache}, {\em A
  {L}og-{E}uclidean framework for statistics on diffeomorphisms}, in Proc
  Medical Image Computing and Computer-Assisted Intervention, vol.~LNCS 4190,
  2006, pp.~924--931.

\bibitem{Ashburner:2007a}
{\sc J.~Ashburner}, {\em A fast diffeomorphic image registration algorithm},
  NeuroImage, 38 (2007), pp.~95--113.

\bibitem{Ashburner:2011a}
{\sc J.~Ashburner and K.~J. Friston}, {\em Diffeomorphic registration using
  geodesic shooting and {G}auss-{N}ewton optimisation}, NeuroImage, 55 (2011),
  pp.~954--967.

\bibitem{Avants:2008a}
{\sc B.~B. Avants, C.~L. Epstein, M.~Brossman, and J.~C. Gee}, {\em Symmetric
  diffeomorphic image registration with cross-correlation: {E}valuating
  automated labeling of elderly and neurodegenerative brain}, Medical Image
  Analysis, 12 (2008), pp.~26--41.

\bibitem{Avants:2011a}
{\sc B.~B. Avants, N.~J. Tustison, G.~Song, P.~A. Cook, A.~Klein, and J.~C.
  Gee}, {\em A reproducible evaluation of {ANTs} similarity metric performance
  in brain image registration}, NeuroImage, 54 (2011), pp.~2033--2044.

\bibitem{Azencott:2010a}
{\sc R.~Azencott, R.~Glowinski, J.~He, A.~Jajoo, Y.~Li, A.~Martynenko, R.~H.~W.
  Hoppe, S.~Benzekry, and S.~H. Little}, {\em Diffeomorphic matching and
  dynamic deformable surfaces in {3D} medical imaging}, Computational Methods
  in Applied Mathematics, 10 (2010), pp.~235--274.

\bibitem{Balakrishnan:2019a}
{\sc G.~Balakrishnan, A.~Zhao, M.~R. Sabuncu, J.~Guttag, and A.~V. Dalca}, {\em
  {V}oxel{M}orph: {A} learning framework for deformable medical image
  registration}, IEEE Transactions on Medical Imaging,  (2019).
\newblock (in press) DOI: 10.1109/TMI.2019.2897538.

\bibitem{Balay:2016a}
{\sc S.~Balay, S.~Abhyankar, M.~F. Adams, J.~Brown, P.~Brune, K.~Buschelman,
  L.~Dalcin, V.~Eijkhout, W.~D. Gropp, D.~Kaushik, M.~G. Knepley, L.~C.
  McInnes, K.~Rupp, B.~F. Smith, S.~Zampini, and H.~Zhang}, {\em {PETS}c users
  manual}, Tech. Rep. ANL-95/11 - Revision 3.7, Argonne National Laboratory,
  2016.

\bibitem{Barbu:2016a}
{\sc V.~Barbu and G.~Marinoschi}, {\em An optimal control approach to the
  optical flow problem}, Systems \& Control Letters, 87 (2016), pp.~1--9.

\bibitem{Beg:2005a}
{\sc M.~F. Beg, M.~I. Miller, A.~Trouv\'e, and L.~Younes}, {\em Computing large
  deformation metric mappings via geodesic flows of diffeomorphisms},
  International Journal of Computer Vision, 61 (2005), pp.~139--157.

\bibitem{Biros:2005parallel1}
{\sc G.~Biros and O.~Ghattas}, {\em Parallel lagrange--newton--krylov--schur
  methods for pde-constrained optimization. part i: The krylov--schur solver},
  SIAM Journal on Scientific Computing, 27 (2005), pp.~687--713.

\bibitem{Biros:2005parallel2}
\leavevmode\vrule height 2pt depth -1.6pt width 23pt, {\em Parallel
  lagrange--newton--krylov--schur methods for pde-constrained optimization.
  part ii: The lagrange--newton solver and its application to optimal control
  of steady viscous flows}, SIAM Journal on Scientific Computing, 27 (2005),
  pp.~714--739.

\bibitem{Boggs:1995a}
{\sc P.~T. Boggs and J.~W. Tolle}, {\em Sequential quadratic programming}, Acta
  Numerica, 4 (1995), pp.~1--51.

\bibitem{Bone:2018a}
{\sc A.~Bone, O.~Colliot, and S.~Durrleman}, {\em Learning distributions of
  shape trajectories from longitudinal datasets: {A} hierarchical model on a
  manifold of diffeomorphisms}, arXiv e-prints,  (2019).

\bibitem{Bone:2018b}
{\sc A.~Bone, M.~Louis, B.~Martin, and S.~Durrleman}, {\em Deformetrica 4: {A}n
  open-source software for statistical shape analysis}, in Proc International
  Workshop on Shape in Medical Imaging, vol.~LNCS 11167, 2018, pp.~3--13.

\bibitem{Borzi:2002a}
{\sc A.~Borz\`{i}, K.~Ito, and K.~Kunisch}, {\em Optimal control formulation
  for determining optical flow}, SIAM Journal on Scientific Computing, 24
  (2002), pp.~818--847.

\bibitem{Budelmann:2019a}
{\sc D.~Budelmann, L.~Koenig, N.~Papenberg, and J.~Lellmann}, {\em
  Fully-deformable {3D} image registration in two seconds}, in Bildverarbeitung
  f\"ur die Medizin, 2019, pp.~302--307.

\bibitem{Burger:2013a}
{\sc M.~Burger, J.~Modersitzki, and L.~Ruthotto}, {\em A hyperelastic
  regularization energy for image registration}, SIAM Journal on Scientific
  Computing, 35 (2013), pp.~B132--B148.

\bibitem{Champagnat2012a}
{\sc F.~Champagnat and Y.~Le~Sant}, {\em Efficient cubic {B}-spline image
  interpolation on a {GPU}}, Journal of Graphics Tools, 16 (2012),
  pp.~218--232.

\bibitem{Chen:2011a}
{\sc K.~Chen and D.~A. Lorenz}, {\em Image sequence interpolation using optimal
  control}, Journal of Mathematical Imaging and Vision, 41 (2011),
  pp.~222--238.

\bibitem{Christensen:2006a}
{\sc G.~E. Christensen, X.~Geng, J.~G. Kuhl, J.~Bruss, T.~J. Grabowski, I.~A.
  Pirwani, M.~W. Vannier, J.~S. Allen, and H.~Damasio}, {\em Introduction to
  the non-rigid image registration evaluation project}, in Proc Biomedical
  Image Registration, vol.~LNCS 4057, 2006, pp.~128--135.

\bibitem{Courty:2008a}
{\sc N.~Courty and P.~Hellier}, {\em Accelerating {3D} non-rigid registration
  using graphics hardware}, International Journal of Image and Graphics, 8
  (2008), pp.~81--98.

\bibitem{Dembo:1982a}
{\sc R.~S. Dembo, S.~C. Eisenstat, and T.~Steihaug}, {\em Inexact {N}ewton
  methods}, SIAM Journal on Numerical Analysis, 19 (1982), pp.~400--408.

\bibitem{Dupuis:1998a}
{\sc P.~Dupuis, U.~Gernander, and M.~I. Miller}, {\em Variational problems on
  flows of diffeomorphisms for image matching}, Quarterly of Applied
  Mathematics, 56 (1998), pp.~587--600.

\bibitem{deformetrica-git}
{\sc A.~S. Durrleman, A.~Bone, M.~Louis, B.~Martin, P.~Gori, A.~Routier,
  M.~Bacci, A.~Fougier, B.~Charlier, J.~Glaunes, J.~Fishbaugh, M.~Prastawa,
  M.~Diaz, and C.~Doucet}, {\em deformetrica [commit: v4.0.0-390-ged9c1f9;
  libraries: python3.6; cuda9.2.88]}, 2019.

\bibitem{Durrleman:2014a}
{\sc S.~Durrleman, M.~Prastawa, N.~Charon, J.~R. Korenberg, S.~Joshi, G.~Gerig,
  and A.~Trouve}, {\em Morphometry of anatomical shape complexes with dense
  deformations and sparse parameters}, NeuroImage, 101 (2014), pp.~35--49.

\bibitem{Eisenstat:1996a}
{\sc S.~C. Eisentat and H.~F. Walker}, {\em Choosing the forcing terms in an
  inexact {N}ewton method}, SIAM Journal on Scientific Computing, 17 (1996),
  pp.~16--32.

\bibitem{Eklund:2013a}
{\sc A.~Eklund, P.~Dufort, D.~Forsberg, and S.~M. LaConte}, {\em Medical image
  processing on the {GPU}--past, present and future}, Medical Image Analysis,
  17 (2013), pp.~1073--1094.

\bibitem{Ellingwood:2016a}
{\sc N.~D. Ellingwood, Y.~Yin, M.~Smith, and C.-L. Lin}, {\em Efficient methods
  for implementation of multi-level nonrigid mass-preserving image registration
  on {GPU}s and multi-threaded {CPU}s}, Computer Methods and Programs in
  Biomedicine, 127 (2016), pp.~290--300.

\bibitem{Fischer:2008a}
{\sc B.~Fischer and J.~Modersitzki}, {\em Ill-posed medicine -- an introduction
  to image registration}, Inverse Problems, 24 (2008), pp.~1--16.

\bibitem{Fishbaugh:2017a}
{\sc J.~Fishbaugh, S.~Durrleman, M.~Prastawa, and G.~Gerig}, {\em Geodesic
  shape regression with multiple geometries and sparse parameters}, Medical
  Image Analysis, 39 (2017), pp.~1--17.

\bibitem{Fluck:2011a}
{\sc O.~Fluck, C.~Vetter, W.~Wein, A.~Kamen, B.~Preim, and R.~Westermann}, {\em
  A survey of medical image registration on graphics hardware}, Computer
  Methods and Programs in Biomedicine, 104 (2011), pp.~e45--e57.

\bibitem{accfft_github}
{\sc A.~Gholami and G.~Biros}, {\em {AccFFT}}, 2017.

\bibitem{accfft-home-page}
{\sc A.~Gholami and G.~Biros}, {\em {AccFFT} home page}, 2017.

\bibitem{Gholami:2017SC}
{\sc A.~Gholami, A.~Mang, K.~Scheufele, C.~Davatzikos, M.~Mehl, and G.~Biros},
  {\em A framework for scalable biophysics-based image analysis}, in Proc
  ACM/IEEE Conference on Supercomputing, 2017, pp.~1--13.

\bibitem{Grzech:2019a}
{\sc D.~Grzech, L.~Folgoc, M.~P. Heinrich, B.~Khanal, J.~Moll, J.~A. Schnabel,
  B.~Glocker, and B.~Kainz}, {\em {FastReg}: {F}ast non-rigid registration via
  accelerated optimisation on the manifold of diffeomorphisms}, arXiv e-prints,
   (2019).

\bibitem{Gu:2009a}
{\sc X.~Gu, H.~Pan, Y.~Liang, R.~Castillo, D.~Yang, D.~Choi, E.~Castillo,
  A.~Majumdar, T.~Guerrero, and S.~B. Jiang}, {\em Implementation and
  evaluation of various demons deformable image registration algorithms on a
  {GPU}}, Physics in Medicine and Biology, 55 (2009), pp.~207--219.

\bibitem{Ha:2011a}
{\sc L.~Ha, J.~Kr\"uger, S.~Joshi, and C.~T. Silva}, {\em Multiscale unbiased
  diffeomorphic atlas construction on multi-{GPU}s}, in CPU Computing Gems
  Emerald Edition, Elsevier Inc, 2011, ch.~48, pp.~771--791.

\bibitem{Ha:2009a}
{\sc L.~K. Ha, J.~Kr\"uger, P.~T. Fletcher, S.~Joshi, and C.~T. Silva}, {\em
  Fast parallel unbiased diffeomorphic atlas construction on multi-graphics
  processing units}, in Proc Eurographics Conference on Parallel Grphics and
  Visualization, 2009, pp.~41--48.

\bibitem{cudafinitedifference-git}
{\sc M.~Harris}, {\em Nvidia developer blog}, 2019.

\bibitem{Hernandez:2009a}
{\sc M.~Hernandez, M.~N. Bossa, and S.~Olmos}, {\em Registration of anatomical
  images using paths of diffeomorphisms parameterized with stationary vector
  field flows}, International Journal of Computer Vision, 85 (2009),
  pp.~291--306.

\bibitem{Hestenes:1952a}
{\sc M.~R. Hestenes and E.~Stiefel}, {\em Methods of conjugate gradients for
  solving linear systems}, Journal of Research of the National Bureau of
  Standards, 49 (1952), pp.~409--436.

\bibitem{Hinze:2009a}
{\sc M.~Hinze, R.~Pinnau, M.~Ulbrich, and S.~Ulbrich}, {\em Optimization with
  {PDE} constraints}, Springer, Berlin, DE, 2009.

\bibitem{Joshi:2005a}
{\sc S.~Joshi, B.~Davis, M.~Jornier, and G.~Gerig}, {\em Unbiased diffeomorphic
  atlas construction for computational anatomy}, NeuroImage, 23 (2005),
  pp.~S151--S160.

\bibitem{Klein:2010a}
{\sc S.~Klein, M.~Staring, K.~Murphy, M.~A. Viergever, and J.~P.~W. Pluim},
  {\em {ELASTIX}: {A} tollbox for intensity-based medical image registration},
  Medical Imaging, IEEE Transactions on, 29 (2010), pp.~196--205.

\bibitem{Koenig:2018a}
{\sc L.~Koenig, J.~Ruehaak, A.~Derksen, and J.~Lellmann}, {\em A matrix-free
  approach to parallel and memory-efficient deformable image registration},
  SIAM Journal on Scientific Computing, 40 (2018), pp.~B858--B888.

\bibitem{Krebs:2019a}
{\sc J.~Krebs, H.~Delingette, B.~Mailh\'e, N.~Ayache, and T.~Mansi}, {\em
  Learning a probabilistic model for diffeomorphic registration}, IEEE
  Transactions on Medical Imaging,  (2019).
\newblock (in press) DOI: 10.1109/TMI.2019.2897112.

\bibitem{Lorenzi:2013b}
{\sc M.~Lorenzi, N.~Ayache, G.~B. Frisoni, and X.~Pennec}, {\em {LCC}-demons: a
  robust and accurate symmetric diffeomorphic registration algorithm},
  NeuroImage, 81 (2013), pp.~470--483.

\bibitem{Lorenzi:2013a}
{\sc M.~Lorenzi and X.~Pennec}, {\em Geodesics, parallel transport and
  one-parameter subgroups for diffeomorphic image registration}, International
  Journal of Computer Vision, 105 (2013), pp.~111--127.

\bibitem{Mang:2015NK}
{\sc A.~Mang and G.~Biros}, {\em An inexact {N}ewton--{K}rylov algorithm for
  constrained diffeomorphic image registration}, SIAM Journal on Imaging
  Sciences, 8 (2015), pp.~1030--1069.

\bibitem{Mang:2016H1}
\leavevmode\vrule height 2pt depth -1.6pt width 23pt, {\em Constrained
  {$H^1$}-regularization schemes for diffeomorphic image registration}, SIAM
  Journal on Imaging Sciences, 9 (2016), pp.~1154--1194.

\bibitem{Mang:2017SL}
\leavevmode\vrule height 2pt depth -1.6pt width 23pt, {\em A
  {S}emi-{L}agrangian two-level preconditioned {N}ewton--{K}rylov solver for
  constrained diffeomorphic image registration}, SIAM Journal on Scientific
  Computing, 39 (2017), pp.~B1064--B1101.

\bibitem{claire-web}
{\sc A.~Mang and G.~Biros}, {\em Constrained large deformation diffeomorphic
  image registration ({CLAIRE})}, 2019.
\newblock [Commit: v0.07-131-gbb7619e].

\bibitem{Mang:2016SC}
{\sc A.~Mang, A.~Gholami, and G.~Biros}, {\em Distributed-memory
  large-deformation diffeomorphic {3D} image registration}, in Proc ACM/IEEE
  Conference on Supercomputing, 2016.

\bibitem{Mang:2018CLAIRE}
{\sc A.~Mang, A.~Gholami, C.~Davatzikos, and G.~Biros}, {\em {CLAIRE:} a
  distributed-memory solver for constrained large deformation diffeomorphic
  image registration}, SIAM Journal on Scientific Computing, 41 (2019),
  pp.~C548--C584.

\bibitem{Miller:2001a}
{\sc M.~I. Miller and L.~Younes}, {\em Group actions, homeomorphism, and
  matching: {A} general framework}, International Journal of Computer Vision,
  41 (2001), pp.~61--81.

\bibitem{Modat:2010a}
{\sc M.~Modat, G.~R. Ridgway, Z.~A. Taylor, M.~Lehmann, J.~Barnes, D.~J.
  Hawkes, N.~C. Fox, and S.~Ourselin}, {\em Fast free-form deformation using
  graphics processing units}, Computer Methods and Programs in Biomedicine, 98
  (2010), pp.~278--284.

\bibitem{Modersitzki:2004a}
{\sc J.~Modersitzki}, {\em Numerical methods for image registration}, Oxford
  University Press, New York, 2004.

\bibitem{Modersitzki:2009a}
\leavevmode\vrule height 2pt depth -1.6pt width 23pt, {\em {FAIR}: Flexible
  algorithms for image registration}, SIAM, Philadelphia, Pennsylvania, US,
  2009.

\bibitem{Nocedal:2006a}
{\sc J.~Nocedal and S.~J. Wright}, {\em Numerical Optimization}, Springer, New
  York, New York, US, 2006.

\bibitem{Nvidia2007b}
{\sc Nvidia}, {\em {CUDA CUFFT Library}}, 2007.

\bibitem{pytorch-git}
{\sc A.~Paszke, S.~Gross, S.~Chintala, and G.~Chanan}, {\em Tensors and dynamic
  neural networks in python with strong {GPU} acceleration}, 2019.

\bibitem{Polzin:2016a}
{\sc T.~Polzin, M.~Niethammer, M.~P. Heinrich, H.~Handels, and J.~Modersitzki},
  {\em Memory efficient {LDDMM} for lung {CT}}, in Proc Medical Image Computing
  and Computer-Assisted Intervention, vol.~LNCS 9902, 2016, pp.~28--36.

\bibitem{pyca-git}
{\sc J.~S. Preston}, {\em Python for computational anatomy}, 2019.
\newblock [Commit: v0.01-434-gf31ab43; Libraries: ITK4.13.2; boost1.69;
  FFTW3.3.6-pl2; python2.7; CUDA9.2.88].

\bibitem{Rueckert:1999a}
{\sc D.~Rueckert, L.~I. Sonoda, C.~Hayes, D.~L.~G. Hill, M.~O. Leach, and D.~J.
  Hawkes}, {\em Non-rigid registration using free-form deformations:
  Application to breast {MR} images}, Medical Imaging, IEEE Transactions on, 18
  (1999), pp.~712--721.

\bibitem{cubictextureinterp-git}
{\sc D.~Ruijters}, {\em {GPU} accelerated pre-filtered cubic {B}-spline
  interpolation using {CUDA}}, 2019.

\bibitem{Ruijters:2008a}
{\sc D.~Ruijters, B.~ter Haar~Romeny, and P.~Suetens}, {\em Efficient gpu-based
  texture interpolation using uniform b-splines}, Journal of Graphics Tools, 13
  (2008), pp.~61--69.

\bibitem{Thevenaz:2012a}
{\sc D.~Ruijters and P.~Th{\'e}venaz}, {\em {GPU} prefilter for accurate cubic
  {B}-spline interpolation}, The Computer Journal, 55 (2012), pp.~15--20.

\bibitem{Shackleford:2010a}
{\sc J.~Shackleford, N.~Kandasamy, and G.~Sharp}, {\em On developing {B}-spline
  registration algorithms for multi-core processors}, Physics in Medicine and
  Biology, 55 (2010), pp.~6329--6351.

\bibitem{Shamonin:2014a}
{\sc D.~P. Shamonin, E.~E. Bron, B.~P.~F. Lelieveldt, M.~Smits, S.~Klein, and
  M.~Staring}, {\em Fast parallel image registration on {CPU} and {GPU} for
  diagnostic classification of {A}lzheimer's disease}, Frontiers in
  Neuroinformatics, 7 (2014), pp.~1--15.

\bibitem{Shams:2010a}
{\sc R.~Shams, P.~Sadeghi, R.~A. Kennedy, and R.~I. Hartley}, {\em A survey of
  medical image registration on multicore and the {GPU}}, Signal Processing
  Magazine, IEEE, 27 (2010), pp.~50--60.

\bibitem{gpugems}
{\sc C.~Sigg and M.~Hadwiger}, {\em Fast third-order texture filtering},
  (2005), pp.~313--329.

\bibitem{Sommer:2011a}
{\sc S.~Sommer}, {\em Accelerating multi-scale flows for {LDDKBM} diffeomorphic
  registration}, in Proc IEEE International Conference on Computer Visions
  Workshops, 2011, pp.~499--505.

\bibitem{Sotiras:2013a}
{\sc A.~Sotiras, C.~Davatzikos, and N.~Paragios}, {\em Deformable medical image
  registration: {A} survey}, Medical Imaging, IEEE Transactions on, 32 (2013),
  pp.~1153--1190.

\bibitem{Trouve:1998a}
{\sc A.~Trouv\'e}, {\em Diffeomorphism groups and pattern matching in image
  analysis}, International Journal of Computer Vision, 28 (1998), pp.~213--221.

\bibitem{ValeroLara:2013a}
{\sc P.~Valero-Lara}, {\em A {GPU} approach for accelerating {3D} deformable
  registration ({DARTEL}) on brain biomedical images}, in Proc European MPI
  Users' Group Meeting, 2013, pp.~187--192.

\bibitem{ValeroLara:2014a}
{\sc P.~Valero-Lara}, {\em Multi-{GPU} acceleration of {DARTEL} (early
  detection of {A}lzheimer)}, in Proc IEEE International Conference on Cluster
  Computing, 2014, pp.~346--354.

\bibitem{Vercauteren:2009a}
{\sc T.~Vercauteren, X.~Pennec, A.~Perchant, and N.~Ayache}, {\em Diffeomorphic
  demons: {E}fficient non-parametric image registration}, NeuroImage, 45
  (2009), pp.~S61--S72.

\bibitem{Vialard:2012a}
{\sc F.-X. Vialard, L.~Risser, D.~Rueckert, and C.~J. Cotter}, {\em
  Diffeomorphic {3D} image registration via geodesic shooting using an
  efficient adjoint calculation}, International Journal of Computer Vision, 97
  (2012), pp.~229--241.

\bibitem{roofline}
{\sc S.~Williams, A.~Waterman, and D.~Patterson}, {\em Roofline: An insightful
  visual performance model for multicore architectures}, Commun. ACM, 52
  (2009), pp.~65--76.

\bibitem{Yang:2016a}
{\sc X.~Yang, R.~Kwitt, and M.~Niethammer}, {\em Fast predictive image
  registration}, in Proc International Workshop on Deep Learning in Medical
  Image Analysis, 48-57, ed., vol.~LNCS 10008, 2016, pp.~48--57.

\bibitem{Yang:2017a}
{\sc X.~Yang, R.~Kwitt, M.~Styner, and M.~Niethammer}, {\em Quicksilver: {F}ast
  predictive image registration---{A} deep learning approach}, NeuroImage, 158
  (2017), pp.~378--396.

\bibitem{Younes:2007a}
{\sc L.~Younes}, {\em Jacobi fields in groups of diffeomorphisms and
  applications}, Quarterly of Applied Mathematics, 650 (2007), pp.~113--134.

\bibitem{Younes:2010a}
\leavevmode\vrule height 2pt depth -1.6pt width 23pt, {\em Shapes and
  diffeomorphisms}, Springer, 2010.

\bibitem{Younes:2009a}
{\sc L.~Younes, F.~Arrate, and M.~I. Miller}, {\em Evolutions equations in
  computational anatomy}, NeuroImage, 45 (2009), pp.~S40--S50.

\bibitem{Zhang:2015b}
{\sc M.~Zhang and P.~Fletcher}, {\em Finite-dimensional lie algebras for fast
  diffeomorphic image registration}, in Proc Information Processing in Medical
  Imaging, Springer International Publishing, 2015, pp.~249--260.

\bibitem{Zhang:2018a}
{\sc M.~Zhang and P.~T. Fletcher}, {\em Fast diffeomorphic image registration
  via {F}ourier-approximated {L}ie algebras}, International Journal of Computer
  Vision,  (2018), pp.~1--13.

\end{thebibliography}
